\documentclass[11pt]{elsarticle}
\biboptions{sort&compress,numbers}
\usepackage{graphicx}
\usepackage{amsmath,amssymb,amsthm,bm,stmaryrd}
\usepackage{cancel}
\usepackage{float}
\usepackage{miller}
\usepackage{subcaption}
\usepackage[colorinlistoftodos]{todonotes}
\usepackage[margin=1in]{geometry}
\usepackage{esint}
\usepackage{enumitem}
\usepackage[most]{tcolorbox}
\usepackage{algorithm}
\usepackage{algpseudocode}
\usepackage{soul}
\usepackage[hidelinks]{hyperref}
\usepackage{tabularx}
\usepackage{booktabs}

\begin{document}
    \begin{frontmatter}
        \title{One-dimensional moving window atomistic framework to model long-time shock wave propagation}
        \author[auburn]{Alexander Davis}
        \author[auburn]{Vinamra Agrawal\corref{cor1}}
        \ead{vinagr@auburn.edu}
        \cortext[cor1]{Corresponding author}
        \address[auburn]{Aerospace Engineering Department, Auburn University, Auburn, AL USA}
        \begin{abstract}
            We develop a long-time moving window framework using Molecular Dynamics (MD) to model shock wave propagation through a one-dimensional chain of atoms. 
            The domain is divided into a purely atomistic ``window'' region containing the shock wave flanked by boundary or ``continuum'' regions on either end. 
            The dynamics of the window atoms are governed by classic MD equations of motion while continuum shock conditions are applied to the continuum atoms.
            Spurious wave reflections are removed by employing a damping band method using the Langevin thermostat which is applied locally to the atoms in each continuum region.
            We then implement ideas of control volume on the MD framework such that a moving window follows a propagating shock. 
            The moving window effect is achieved by adding/removing atoms to/from the window and boundary regions, and thus the shock wave front is focused at the center of the window region indefinitely. 
            As a result, the required domain size is very small which significantly reduces the computational complexity of the simulations and allows us to model shock wave propagation much longer than conventional non-equilibrium MD (NEMD) shock techniques.
            We simulate the shock through a one-dimensional chain of copper atoms using either the Lennard-Jones, modified Morse, or Embedded Atom Model (EAM) interatomic potential. 
            We first perform verification studies to ensure proper implementation of the thermostat, potential functions, and damping band method, respectively. 
            Next, we track the propagating shock and compare the actual shock velocity and average particle velocity to their corresponding analytical input values. 
            From these comparisons, we optimize the linear shock Hugoniot relation for the given ``lattice" orientation and compare these results to those in literature.
            When incorporated into the linear shock equation, these new Hugoniot parameters are shown to produce a stationary shock wave front.
            Finally, we perform one-dimensional moving window simulations of an unsteady, structured shock up to a few nanoseconds and characterize the increase in the shock front's width.    
        \end{abstract}
    \end{frontmatter}

    \section{Introduction}
        The propagation of shock waves in materials is an important scientific phenomenon that has been widely studied for many decades; see, for example, \cite{meyers1994dynamic, davison2008fundamentals} and the references therein. 
        It is thus well known that the shock response of a material at any length scale is inextricably linked to its response at lower length scales.
        At the macroscopic scale, shock waves can lead to damage, plastic deformation, and fracture of the material.
        At the micro- and meso-scale, shock waves can interact with the microstructure causing complex behavior including scattering, grain rotations, pore collapse, phase transformations, dislocation and void generation, and grain crushing \cite{RustyGray2012,Fensin2014,Bingert2014}.
        
        Because shock phenomena at the lower length scales influence behavior at the higher length scales, the shock response of materials has been modeled extensively at the atomistic level using non-equilibrium Molecular Dynamics (NEMD) simulations since the 1960s \cite{holian1995atomistic}.
        In the past two decades, these NEMD shock simulations have been expanded to very large-scale domains and used to model increasingly complex events \cite{holian1998plasticity, germann2000orientation,germann2004dislocation,Bringa2004,bringa2006shock,Srinivasan2007,Norman2008,bringa2010void,Fensin2014Effect,perriot2014evolution,bisht2019investigation}.
        These simulations typically involve several millions of atoms $(\sim O(10^5 - 10^9))$ subjected to flyer-plate loading scenarios.
        Such simulations have been used to study void nucleation \cite{bringa2010void}, dislocation generation \cite{Tramontina2017Simulation}, twinning \cite{Higginbotham2013Molecular} and even shock induced spallation \cite{Fensin2013Effect,Fensin2014,Fensin2014Effect}.
        However, because of limited domain sizes, NEMD shock methods can suffer from wave reflections off the domain boundary. 
        Such incidents lead to transient effects and drastically reduce the total simulation time.
        Additionally, NEMD shock simulations typically result in unrealistic strain rates $(10^{10}-10^{12}\, s^{-1})$ \cite{meyers1994dynamic}.
        Such strain rates are rare and orders of magnitude higher than those observed in experiments and practical scenarios $(10^6-10^8\, s^{-1})$.
        
        To overcome some of the drawbacks with NEMD methods, alternative atomistic frameworks have been developed to model shock waves.
        Two of these methods are the uniaxial Hugoniostat \cite{maillet2000uniaxial,maillet2002uniaxial,ravelo2004constant,bedrov2009shock} and the multiscale shock technique (MSST) \cite{Reed2003,reed2006analysis}.
        The uniaxial Hugoniostat method compresses the crystal instantaneously to the final shocked volume and then couples the system to a thermostat which guarantees that the final Hugoniot state is reached. 
        This framework has been shown to reproduce defects generated by the shock wave and is an order of magnitude less expensive than classic NEMD simulations. 
        However, the Hugoniostat is an equilibrium method which merely reproduces the final shocked state in the domain, so the study of a shock's steadiness and structure as well as its interaction with other shocks and defects is limited. 
        On the other hand, MSST performs long-time shock simulations on small atomistic domains for a much lower computational cost than conventional NEMD shock simulations. 
        MSST can also simulate multiple shock waves in the atomistic domain.
        While this technique permits the shock to be controlled based on prescribed continuum constraints, MSST does not allow information such as defects and heat to be transferred between the atomistic and continuum regions. 
        As a result, the scalability of MSST to a fully-coupled atomisitic/continuum scheme is restricted.

        While modern atomistic methods have been very successful in modeling shock waves and characterizing how defects influence shock propagation, such schemes still suffer from various limitations as described above. 
        A concurrent multiscale scheme is needed that would allow the atomistic region containing the shock wave to follow the shock. 
        This would require simultaneously refining the continuum region as well as coarsening the atomistic region at the speed of the propagating shock wave.
        Such a framework would permit microstructural defects and the resulting scattered elastic waves to consistently cross the interface to the continuum region.
        This paper presents a first step towards that concurrent atomistic/continuum scheme by developing a moving window atomistic framework to follow a propagating shock wave through a material.

        In this work, we develop a long-time moving window atomistic framework using Molecular Dynamics (MD) to model shock wave propagation through a one-dimensional chain of atoms.
        This framework employs techniques similar to those applied in the uniaxial ``Hugoniostat" method by using the planar shock jump conditions and Hugoniot equation of state (EOS) to study the classic Riemann problem of a single propagating shock.
        The domain is divided into an inner ``window" region containing the shock wave flanked by a ``boundary" region on both sides.
        This boundary region is modeled after the ``damping band" method presented in \cite{qu2005finite} which applies a Langevin thermostat locally to continuum atoms in a ``stadium" fashion and linearly increases the damping coefficient across the thermostatted region. 
        Our method differs from \cite{qu2005finite} because it is purely atomistic and thus does not further couple the stadium boundary region to an outer continuum region.
        The damping band regions absorb and abate any impinging waves thus largely eliminating transient wave reflections. 
        The atoms in the boundary regions are governed by continuum shock equations and act as boundary conditions for the window atoms.
        
        The motion of the domain is achieved by consistently adding and removing atoms to and from the boundary and window regions.
        This moving window technique is similar in principle to moving boundary conditions used in works such as \cite{holland1998ideal} and \cite{selinger2000dynamic} to model dynamic crack propagation, while incorporation of a shock wave into the moving window framework is inspired from \cite{Zhakhovskii1997} and \cite{zhakhovsky2011two}.
        Ordinarily, the simulation time of a shock propagating in an atomistic domain would be vastly limited due to wave reflections off the boundary.
        The moving window formulation allows us to model the propagating shock much longer than conventional NEMD shock simulations by focusing the shock front at the center of the window region for the entire simulation.
        We emphasize that the framework, in its current state, is not a truly concurrent atomistic/continuum scheme in that it lacks a continuum region with finite element-type mesh points.
        The scope of this work is therefore limited to ensuring that the small domain formulation can follow a shock wave for a long time without artificial wave generation and reflection.

        Next, we use this method to calculate the shock Hugoniot relation of single-crystal copper along the [110] close packed lattice direction.
        Much work has been done on shock kinetic relations and the linear Hugoniot relationship between shock velocity and particle speed.
        This includes extensive experimental calibration of the linear relation \cite{marsh1980lasl,mitchell1991equation}, as well as theoretical investigations into the origins of the shock kinetic relation \cite{knowles2002relation}.
        Additionally, many computational studies have been performed which use MD to measure the shock Hugoniot along different orientations of an FCC copper lattice \cite{germann2000orientation,Bringa2004,lin2014effects,neogi2017shock}.
        While the MD work has shown large anisotropic behavior for shock propagation along different crystal directions of single-crystal copper, experimental studies have shown no crystal orientation dependence of the shock velocity vs. particle velocity Hugoniot curve \cite{chau2010shock}.
        
        We start with the experimentally known linear law for polycrystalline bulk copper \cite{marsh1980lasl,mitchell1991equation,davison2008fundamentals} and perform moving window simulations using shock velocities obtained from the known kinetic relation. 
        Next, we provide modifications to the linear law using the one-dimensional chain of atoms to ensure a stationary shock wave. 
        This new relation between shock velocity and particle velocity is defined as the shock Hugoniot curve along the [110] direction of a single-crystal copper lattice and compared to other MD studies. 
        Finally, we use this optimized Hugoniot data along with the moving window to follow a structured shock up to a few nanoseconds and characterize the shock front's width.
        The shock's width is observed to increase with time which implies that the shock wave is unsteady.
        This is consistent with the findings of other early MD studies which used a one-dimensional chain of atoms to model shock wave propagation \cite{tsai1966shock,duvall1969steady,holian1978molecular,straub1979molecular,holian1995atomistic}.

        The paper is organized as follows. 
        Section \ref{ProblemStatement} outlines the classic Riemann problem of a single propagating shock wave with constant states across it and relates a one-dimensional chain of atoms to a bulk FCC copper lattice. 
        Section \ref{AtomFram} presents the computational components of the framework as well as describes the application of the jump parameters and origination of the shock.
        Section \ref{LangevinThermostat} discusses the Langevin ``damping band" technique and shows how it is implemented in the atomistic domain.
        Section \ref{MovingWindowSec} describes the moving window formulation and relates it to previous works.
        Section \ref{Verification} presents detailed verifications of every component of the framework (some of which is found in the appendix).
        Section \ref{ResultsEOS} uses the moving window framework to follow a propagating shock and calculate the shock velocity vs. particle velocity Hugoniot along the [110] direction of a single-crystal copper lattice. 
        This shock Hugoniot is shown to be in very good agreement with the results of other MD studies.  
        Finally, Section \ref{ResultsShockWidth} performs long-time simulations of the shock up to a few nanoseconds and characterizes the shock's structure while relating the results to early 1D shock studies. 
        
    \section{Problem Statement} \label{ProblemStatement}

        We model a shock wave using the conservation of momentum, continuity equation, Hugoniot EOS, and a thermodynamic relationship. 
        At the continuum level, the one-dimensional shock equations are given by the following jump conditions \cite{davison2008fundamentals}:
        \begin{align}
            \llbracket \sigma \rrbracket + \rho U_s \llbracket v \rrbracket &= 0 \label{MomentumEquation} \\
            \llbracket v \rrbracket + U_s \llbracket \epsilon \rrbracket &= 0\label{MassEquation}
        \end{align}
        where $\sigma$, $\epsilon$, $\rho$ and $v$ denote stress, strain, density, and particle velocity respectively. 
        The speed of the propagating shock front is $U_s$, and $\llbracket\cdot\rrbracket$ denotes the change in the given quantity across the shock.
        The shock jump equations are supplemented by an empirically observed linear relation between shock velocity and particle velocity,
        \begin{equation}
            U_s = C_0 + S\llbracket v\rrbracket. \label{eq:LinearLaw}
        \end{equation}
        Here, \textit{S} is a dimensionless, empirical parameter representing the slope of the shock velocity vs. particle velocity Hugoniot curve, and \textit{$C_{0}$} is the sound velocity in the material at zero stress.
        Equations (\ref{MomentumEquation}), (\ref{MassEquation}), and (\ref{eq:LinearLaw}) lead to the standard Hugoniot stress-strain relationship given by 
        \begin{equation}
            \sigma = \frac{\rho C_0^2 \llbracket \epsilon \rrbracket}{(1 + S \llbracket \epsilon \rrbracket)^2} \label{eq:HugoniotEOS}
        \end{equation}
        where compression stress and strain are considered positive. 
        The Hugoniot stress-strain relationship forms the basis of modern equations of state.
        In this work, we use the Hugoniot EOS (\ref{eq:HugoniotEOS}) for the sake of simplicity and to quickly compute shocked states.
        Finally, the temperature change across the shock can be calculated by solving the following differential equation:
        \begin{equation}
            C_{V} \left(\frac{dT}{d \epsilon} \right)_{H} - \frac{\gamma T C_{V}}{1 - \epsilon} = \frac{\epsilon}{2} \left(\frac{d \sigma}{d \epsilon} \right)_{H} - \frac{\sigma}{2} \label{eq:ShockHeatEquation}
        \end{equation}
        where $C_V$ is the volumetric specific heat capacity, and $\gamma$ is the Mie-Gruneisen parameter for the material (2.0 for copper \cite{davison2008fundamentals}). 

\begin{figure}[H]
            \centering
            \includegraphics[width=0.4\textwidth]{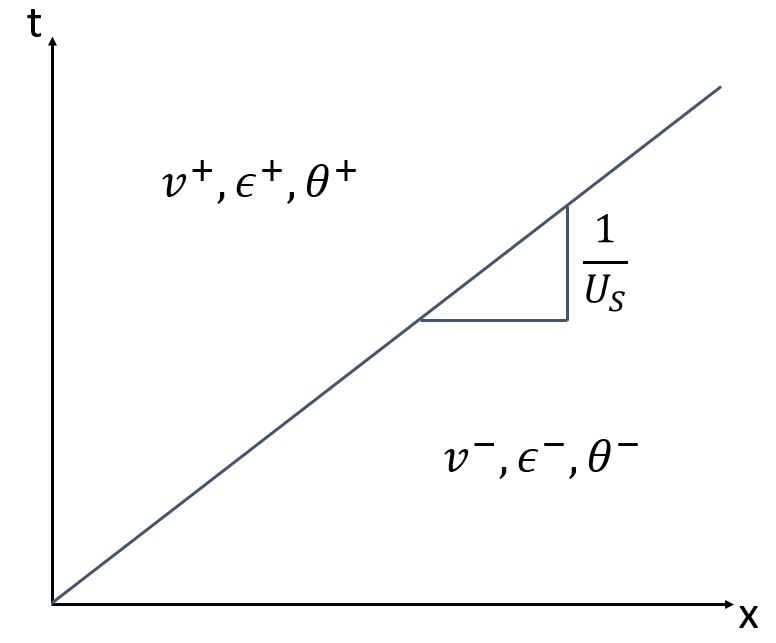}
            \caption{\textit{Riemann problem of a shock wave with constant states ahead of and behind the shock front.}}
            \label{fig:Riemann Shock}
        \end{figure}

        The state of the material is described by the state variables $(v,\epsilon,\theta)$ on either side of the shock, where $\theta$ denotes temperature.
        Given a shock velocity $U_s$ and state $(v^-,\epsilon^-,\theta^-)$ of the unshocked material, equations (\ref{MomentumEquation}), (\ref{MassEquation}), (\ref{eq:HugoniotEOS}), and (\ref{eq:ShockHeatEquation}) can be used to compute the state $(v^+,\epsilon^+,\theta^+)$ and stress $\sigma$ of the shocked material.
        In this paper, we present a moving window framework to simulate long-time shock wave propagation using atomistics given continuum shock states ahead of and behind the shock. 
        In this sense, we study the classic Riemann problem of a single shock wave with constant states on either side as shown in Fig. \ref{fig:Riemann Shock}.
        
        We study shock wave propagation through an idealized, one-dimensional, ``close packed" chain of copper (Cu) atoms.
        The lattice points in the one-dimensional chain are chosen to represent copper atoms and hence are each assigned a mass of $63.55$ amu. 
        For the potential functions used in this work (Lennard-Jones, modified Morse, and EAM), the equilibrium spacing between copper atoms in a one-dimensional setting is given as $r_0 = 2.5471$ \AA.
        This corresponds to the spacing between atoms along the close packed direction of a bulk FCC copper lattice. 
        Therefore, we define the idealized, one-dimensional chain of copper atoms as ``close packed."
        In a three-dimensional sense, this framework approximates a planar elastic shock propagating along the $[110]$ direction of a single-crystal copper lattice and corresponds to the condition of uniaxial strain for a continuum along this orientation.
        This statement requires some further elaboration and justification.
        
        In a series of papers in the 60s and 70s exploring shock wave propagation and its structure, it was discovered that while a one-dimensional system can be used to recover a linear Hugoniot relation, the width of the observed 1D shock wave increased linearly \cite{tsai1966shock, duvall1969steady, manvi1969shock, manvi1969finite, holian1978molecular,straub1979molecular}.
        The transition from unsteady to steady waves (with constant width) in three-dimensional lattices for shocks above a critical strength was later found to be due to the ``increase in coupling between vibrational excitations normal and transverse to the direction of shock wave propagation'' \cite{holian1979molecular}.
        Plastic deformation and the associated transverse atomic motion, which is obviously never possible in a one-dimensional chain, was thus found to be the key to steady wave behavior in three-dimensional lattices \cite{straub1980molecular,holian1988modeling,holian1998plasticity}. 
        However, one-dimensional chains can still be used to study weak shocks where plastic deformation is minimal \cite{holian1995atomistic}.
        
        It should be emphasized that the early MD simulations of shock waves in 1D were all performed along the [100] lattice direction.
        More recent computational studies have measured the shock Hugoniot along different orientations of a single-crystal copper lattice and discovered large anisotropic behavior along the various crystal directions \cite{germann2000orientation,Bringa2004,lin2014effects,neogi2017shock}.
        Specifically, \cite{germann2000orientation} found that shocks propagating along the [110] direction exhibited a leading solitary wave train, spreading out as time progressed, which was not observed in the other cases. 
        Along this orientation, the atoms displaced primarily along the shock direction and very little, if at all, in the transverse directions. 
        They concluded that shocks along the [110] direction exhibited effectively one-dimensional behavior comparable to what had been observed in cases of a one-dimensional chain of hard rods \cite{holian1978molecular}.
        This was seen for both the Lennard-Jones and EAM potentials. 
        
        Building upon this, we relate a shock wave propagating through a one-dimensional, ``close packed" chain of copper atoms to a planar elastic shock propagating along the [110] direction of a single-crystal copper lattice.
        We maintain low particle velocities ($< 1.6$ km/s) as well as temperatures below the melting temperature of copper ($1,358$ K \cite{davison2008fundamentals}) to study weak shocks.
        This minimizes plastic behavior such as dislocation generation and void nucleation in the shocked region.
        We note, however, that some plastic behavior may still occur.
        Since the one-dimensional framework is fundamentally incapable of capturing plastic effects (and the associated transverse atomic displacements), we should still expect to observe the shock front increase in thickness.
        This is consistent with our observations in Sec. \ref{ResultsShockWidth}.

        We first run simulations using the experimentally known Hugoniot parameters of polycrystalline bulk copper: $S = 1.49$ and $C_0 = 3.94$ km/sec \cite{marsh1980lasl,mitchell1991equation,meyers1994dynamic}.
        We use these values as an initial guess of the linear shock parameters and later utilize the moving window technique to obtain values which produce a stationary shock wave.  
        This is accomplished by plotting the average shock velocity vs. particle velocity for a number of different input shock speeds and using a linear regression analysis to fit the data.
        The slope of this regression equation is the new $S$ value and the y-intercept is the new $C_0$ value.
        This new relation between shock velocity and particle velocity is then defined as the shock Hugoniot curve along the [110] direction of a single-crystal copper lattice and is shown to be in very good agreement with the results of other MD studies. 
        This analysis is presented in Sec. \ref{ResultsEOS}.

    \section{Description of the Atomistic Framework} \label{AtomFram}
        The one-dimensional framework is implemented using an in-house C++ code.
        We utilize a chain of $N$ lattice points each of mass $m = 63.55$ amu -- the mass of an individual copper atom. 
        The equilibrium spacing between atoms in a monoatomic chain is determined by the potential function, and each potential (Sec. \ref{InteratomicPotentials}) assumes an equilibrium spacing of $r_0 = 2.5471$ \AA\, for copper in a one-dimensional setting.
        This corresponds to the spacing along the [110] direction of a bulk copper lattice, so we characterize our system as an idealized, one-dimensional, ``close packed" chain of copper atoms where $x_i$ denotes the instantaneous position of the $i^{th}$ atom at time $t$.
        
        \subsection{Geometry and boundary conditions} \label{GeomAndBCs}
\begin{figure}[H]
                \centering
                \includegraphics[width=0.6\textwidth]{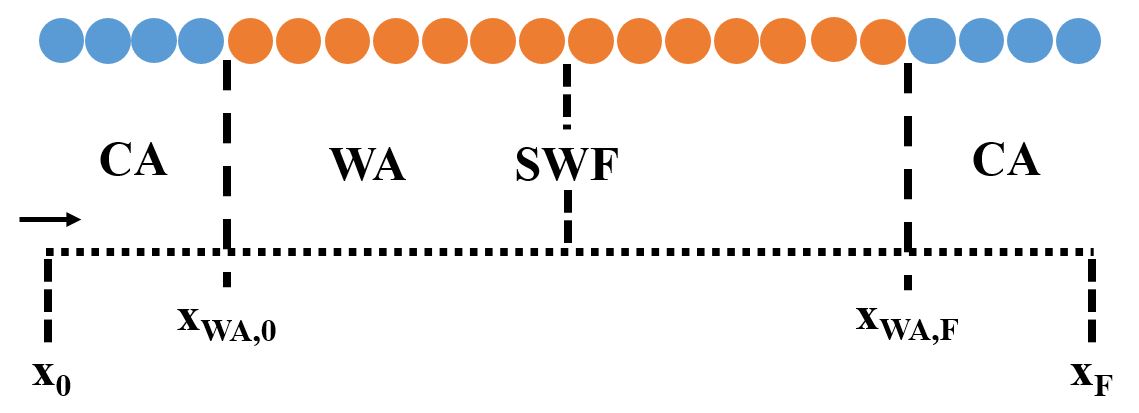}
                \caption{\textit{Schematic of the atomistic framework.}}
                \label{fig:AC Framework}
            \end{figure}

            The atomistic chain is split into three sections as illustrated in Fig. \ref{fig:AC Framework}.
            The outer atoms (in blue) are called \emph{continuum atoms} (CA) while the inner atoms containing the shock wave front (SWF) are called \emph{window atoms} (WA).
            The continuum Reimann states $(v,\epsilon,\theta)$ are imposed on the CA regions using standard algorithms for applying strain, mean particle velocity, and temperature (Langevin thermostat) \cite{tadmor2011modeling}.
            The WA region is governed by classic MD equations.
            To ensure semi-infinite regions on either side of the shock wave, a semi-periodic boundary condition method is employed.
            To achieve this, the continuum atoms at the ends of the chain ($x_{0}$ and $x_{F}$) are made neighbors with the continuum atoms at the WA/CA interfaces ($x_{WA,0}$ and $x_{WA,F}$ respectively). 
            The continuum atoms and window atoms near the WA/CA interfaces interact with each other to ensure smooth information transfer between the two regions.
            
        \subsection{Initialization of the shock wave} \label{InitializationShockWave}
            Typically, shocks are modeled with MD by subjecting a large-scale atomistic domain to flyer-plate (or ``piston") loading scenarios leading to very high strain rates in the shocked material. 
            In contrast, the present formulation initializes the shock wave using techniques inspired from the uniaxial Hugoniostat method \cite{maillet2000uniaxial,maillet2002uniaxial,ravelo2004constant,bedrov2009shock}.
            This formulation uses the Hugoniot conservation relations of mass, momentum, and energy across the shock front (``jump conditions") to simulate the \textit{final state} of the material after the shock has traveled through the domain. 
            Therefore, the Hugoniostat method does not actually simulate the transient processes at the shock front but rather uses the Hugoniot relations as constraints to find the final states on the Hugoniot curve. 
            At time zero, the volume is fixed by allowing the atoms to vibrate around unconstrained lattice sites in the cold, unshocked solid.
            Then, a uniaxial strain is applied from which the final state is reached, and the shock velocity ($U_S$) and particle speed ($v$) can be derived through the jump conditions.
            The final Hugoniot temperature is achieved by coupling the strained system to a feedback thermostat to constrain the internal energy according to the energy conservation equation.
            The Hugoniostat method has been shown to successfully reproduce the Hugoniot curve as well as the defect structures produced by the shock wave. 
            
            The present formulation is similar to the Hugoniostat method in that it uses the Hugoniot jump conditions to characterize the shock wave and then couples a thermostat to a section of the domain, so the prescribed temperature from Eq. (\ref{eq:ShockHeatEquation}) is obtained in the shocked material.
            Our framework differs, however, because the thermostat is applied only to the CA regions, allowing us to simulate the shock and evaluate its structure. 
            To maintain consistency, we always define the SWF to originate at the center of the WA region. 
            All the atoms to the right of the SWF constitute the unshocked material, while all the atoms to the left of the SWF constitute the shocked material.
            For every shock simulation, the unshocked state of the material is specified as follows: ($v^- = 0$ km/s, $\epsilon^- = 0$, $\theta^- = 298$ K).
            It should be noted that the Hugoniot parameters are typically reported for a material with this initial state \cite{davison2008fundamentals}.
            However, the framework will be valid for other unshocked states, provided suitable Hugoniot parameters in Eq. (\ref{eq:LinearLaw}) can be obtained.
            The unshocked (initial) temperature is imposed on the CA region to the right.
            Since the CA region is coupled to the WA region, this ensures that all of the unshocked material maintains a mean temperature of 298 K as shown in \cite{qu2005finite}. 
            
            We then choose a shock wave velocity and use equations (\ref{MassEquation}) and (\ref{eq:LinearLaw}) to obtain the mean particle velocity and strain for the shocked material.
            This mean particle velocity represents a new equilibrium velocity for the atoms in the shocked region, and the integration algorithm is updated accordingly.
            The imposed strain causes the shocked region to compress uniaxially, and the atoms obey the Cauchy-Born rule such that their positions in the chain follow the overall strain of the shocked region.
            The non-zero particle velocity and compressive strain cause the shocked region to reach the final state and produce a forward propagating shock wave starting at the center of the WA region.
            The temperature rise from the shock wave is calculated from Eq. (\ref{eq:ShockHeatEquation}) and imposed in the left CA region which causes the shocked material to maintain this mean temperature \cite{qu2005finite}. 
            The parameters ($v^+$, $\epsilon^+$, $\theta^+$) represent the entire state of the shocked material. 
            It should be noted that the size of CA regions are chosen such that they are far away from the SWF (the non-equilibrium region).
            As discussed in \cite{holian1998plasticity}, the shock velocity vs. particle velocity Hugoniot relation links a given initial equilibrium state to all possible final equilibrium states for planar shock waves.
            Therefore, the CA damping bands are in regions of ``local" equilibrium, ensuring the validity of applying thermostats onto strained sections of the domain \cite{maillet2000uniaxial}.
            
        
        \subsection{Interatomic potentials} \label{InteratomicPotentials}
            In this work, we use the Lennard-Jones (LJ), modified Morse, and Embedded Atom Model (EAM) interatomic potential functions to study the dependence of the Hugoniot parameters on potentials in a one-dimensional setting.
            This analysis is presented in Sec. \ref{NewEOSCalculations}.
            
            The LJ potential only considers nearest-neighbor interactions and is represented most commonly as \cite{jones1924determination,verlet1967computer},
            \begin{equation}
                V(x_{ij}) = 4\epsilon \left[\left(\frac{\sigma}{x_{ij}}\right)^{12}  - \left(\frac{\sigma}{x_{ij}}\right)^6  \right] = \epsilon \left[\left(\frac{r_0}{x_{ij}}\right)^{12}  - 2\left(\frac{r_0}{x_{ij}}\right)^6  \right]
            \end{equation}
            where $\epsilon$ is the depth of the potential well, $\sigma$ is the finite distance at which the inter-particle potential is zero, $x_{ij} = |x_i - x_j|$ is the absolute distance between particle $i$ and $j$, and $r_0$ is the distance at which the potential reaches the minimum. 
            The parameters for copper are given as follows: $\epsilon = 0.4093$ eV and $\sigma = 2.338$ \AA\,  \cite{lv2011molecular}. 
            
            Like LJ, the modified Morse potential \cite{macdonald1981thermodynamic,wen2015interpolation} only considers nearest neighbor interactions. 
            The expression is given by
            \begin{equation}
                V(x_{ij}) = \frac{D_0}{2B-1}\left[e^{-2A \sqrt{B} (x_{ij} - r_0)} - 2Be^{-A (x_{ij} - r_0) / \sqrt{B}}\right].
            \end{equation}
            Here, we use the following parameters for copper: $r_0 = 2.5471$ \AA, $A = 1.1857$ {\AA}$^{-1}$, $D_0 = 0.5869$ eV, and $B = 2.265$ \cite{wen2015interpolation}. 
            MacDonald and MacDonald \cite{macdonald1981thermodynamic} modified the standard Morse potential to improve the agreement with experimental values for the thermal expansion of copper \cite{wen2015interpolation}. 
            
            Finally, the Embedded Atom Model (EAM) potential \cite{foiles1986embedded} is given by the expression
            \begin{equation}\label{EAMEq}
                V(x_{ij}) = F \left(\sum_{i \ne j} \rho \left(x_{ij} \right) \right) + \frac{1}{2} \sum_{i \ne j} \phi \left(x_{ij} \right).
            \end{equation}
            In this case, the total energy of a particular atom is a function of all the atoms within a cutoff radius $r_c = 5.507$ \AA. 
            Here, $\phi$ is a pair-wise potential function, $\rho$ is the contribution to the electron charge density from atom \textit{j} at the location of atom \textit{i}, and \textit{F} is an embedding function that represents the energy required to place atom \textit{i} into the electron cloud \cite{daw1984embedded}. 
            As shown in \cite{daw1993embedded}, the EAM potential works very well for purely metallic systems with no directional bonding and thus provides a robust means of calculating approximate structure and energetics of materials. 
            In our case, we use the EAM potential file produced by Mishin \cite{mishin2001structural}.
            
        \subsection{Integration algorithm} \label{IntegrationAlgorithm}
            To integrate the equations of motion, we utilize the well-known velocity Verlet algorithm \cite{swope1982computer} as seen below for 1D:
            \begin{align}\label{VelocityVerlet}
                x_{i} \left(t + \delta t \right) &= x_i \left(t \right) + v_{i} \left(t \right) \delta t + \frac{f_{i} \left(t \right)}{2m} \delta t^2 \nonumber \\
                v_{i} \left(t + \frac{\delta t}{2} \right) &= v_{i} \left(t \right) + \frac{\delta t}{2} \frac{f_{i} \left (t \right)}{m} \nonumber \\
                f_{i} \left(t + \delta t \right) &= f_{i} \left(x_{i} \left(t + \delta t \right) \right) \nonumber \\
                v_{i} \left(t + \delta t \right) &= v_{i} \left(t + \frac{\delta t}{2}\right) + \frac{\delta t}{2} \frac{f_{i} \left (t  + \delta t \right)}{m}
            \end{align}
            where $x_{i}$, $v_{i}$, and $f_{i}$ denote the position of the $i^{th}$ particle, its velocity, and the net force acting on it respectively. 
            The time step used in the integration algorithm was chosen to be $\delta t = 0.001$ ps $= 1$ fs.
            The velocity Verlet algorithm is adapted in the presence of the Langevin thermostat as explained in Sec. \ref{LangevinThermostat}.

    \section{Langevin thermostat with stadium damping} \label{LangevinThermostat}

        In this work, we use the ``damping band" method developed by \cite{qu2005finite} to incorporate a thermostat into each CA region.
        This method is based on the ``stadium boundary conditions" proposed by \cite{holian1995fracture} along with the Langevin thermostat employed by \cite{holland1999cracks}. 
        Compared to other approaches, the damping band method is more ad hoc and does require some experimentation to fit the optimal parameters (length of the damped region and value of the maximum damping parameter) for a given application. 
        However, it regulates the temperature of the entire CA region; and, at the same time, prevents spurious reflections by absorbing and dampening any artifact waves that impinge on the WA/CA interfaces \cite{qu2005finite, miller2007hybrid}.
        Additionally, this method allows disturbances in particle velocities to propagate through the atomistic domain without being artificially suppressed by a global thermostat since local temperature fluctuations are not felt by the thermostat until phonons arrive at the WA/CA interface \cite{tadmor2013finite}.
            
        The basic idea of damping bands is that the dynamics of the atoms in the CA regions are modified by the Langevin (Brownian) thermostatting algorithm \cite{lepri2003thermal,liu2004introduction}. 
        The Langevin thermostat is a stochastic thermostat which adds a random force to the particle motion along with a damping term, $\zeta$. 
        The one-dimensional equations of motion of the Langevin thermostat for a particle \textit{i} are as follows:
        \begin{align}\label{Langevin}
            f_{i}^{tot} \left(t \right) &= f_{i} \left(t \right) - \zeta m_{i} v_{i} \left(t \right) + \sqrt{\frac{2 k_{B} \theta \zeta m}{\delta t}} \tilde{h_{i}} \left(t \right) \nonumber \\
            \left<\tilde{h}_{i} \left(t \right) \right> &= 0 \nonumber \\
            \left<\tilde{h}_{i, \alpha} \tilde{h}_{i, \beta} \left(t \right) \right> &= \delta_{\alpha \beta}
        \end{align}
        where $\alpha$ and $\beta$ denote Cartesian components, $m$ is the mass of atom \textit{i}, $\delta t$ is the MD time step, $k_{B}$ is Boltzmann's Constant, and $\tilde{h}_{i}$ is a Gaussian random variable with a mean of zero and a variance of one. 
        The continuum atoms are subjected to continuum states $(v^+,\epsilon^+,\theta^+)$ and $(v^-,\epsilon^-,\theta^-)$.
        Since the Langevin thermostat is local in nature, the target temperatures $\theta^+$ and $\theta^-$ are specified for every atom. 
        We adapt the velocity Verlet algorithm in the presence of the Langevin thermostat by performing the discretization used in LAMMPS (Large-scale Atomic/Molecular Massively Parallel Simulator) \cite{schneider1978molecular}:
        \begin{align}
            v_i \left(t + \frac{\delta t}{2} \right) &= v_i(t) - \frac{\delta t}{2} \left(\frac{\nabla_i V(t)}{m} + \zeta v_i(t) \right) + \sqrt{\frac{\delta t k_B T \zeta}{m}}\tilde{h}_i 
            \nonumber \\
            x_i(t + \delta t) &= x_i(t) + v_i \left(t + \frac{\delta t}{2} \right) \delta t
            \nonumber \\
            v_i \left(t + \delta t \right) &= v_i \left(t + \frac{\delta t}{2} \right) - \frac{\delta t}{2} \left(\frac{\nabla_i V(t + \delta t)}{m} + \zeta v_i \left(t + \frac{\delta t}{2} \right) \right) + \sqrt{\frac{\delta t k_B T \zeta}{m}}\tilde{h}_i.
        \end{align}
        We note that because of the Verlet scheme, the time step in each velocity update is now $\frac{\delta t}{2}$ rather than $\delta t$.
        We generate a different random vector for each particle during each velocity update. 
                
        To ensure force matching across the WA and CA regions as well as simultaneously absorb impinging waves effectively and efficiently, we specify the damping factor $\zeta$ to be a function of position relative to the WA/CA interface. 
        To this end, we utilize the equation developed in \cite{qu2005finite} which linearly ramps the damping in the CA regions as the distance from the WA/CA interface increases. 
        This equation is given as follows:
        \begin{equation}\label{LangevinDamping}
            \zeta = \zeta_{0} \left[1 - \frac{d \left(x \right)}{w} \right]
        \end{equation}
        where $\zeta_{0}$ equals the maximum damping of 1/2 the Debye frequency of copper ($\omega_D$), and $w$ is the length of the CA region. 
        Here, \textit{d} is the minimum distance from the atom at position \textit{x} to the end of the chain (either point $x_{0}$ or $x_{F}$).
        For the left and right CA regions, we define $d$ as follows:
        \begin{align}\label{DampingDistance}
            d_{Left} \left(x \right) &= abs \left(x_i - x_0 \right) \nonumber \\
            d_{Right} \left(x \right) &= abs \left(x_i - x_F \right)
        \end{align}
        Hence, for atoms in the CA regions, the damping coefficient varies linearly from zero at the WA/CA interfaces to $\zeta_{0}$ at the ends of the chain. 
        This allows waves to enter the CA region and slowly be absorbed as they propagate to the end of the chain. 
        This reduces spurious wave reflections and artificial waves introduced at the WA/CA interfaces and thus prevents artificial heating in the WA region \cite{qu2005finite, miller2007hybrid}. 
            
         
    \section{Moving Window Formulation} \label{MovingWindowSec}
        Purely atomistic simulations of shock wave propagation suffer from limited domain sizes due to the computational expense associated with modeling large-scale phenomena with just atoms. 
        As a result, the shock cannot travel far before encountering a boundary, and this vastly limits the overall simulation time.
        In the present work, we address this problem by implementing a moving window method which is similar in principle to moving boundary conditions used in \cite{holland1998ideal} and \cite{selinger2000dynamic} to model dynamic crack propagation.
        Incorporation of a shock wave into the moving window framework is inspired from \cite{Zhakhovskii1997} and \cite{zhakhovsky2011two}, where a constant flux of material with a given density and velocity is fed into the simulation window by inserting a plane of atoms to the right boundary at regular time intervals.
        
\begin{figure}[H]
            \centering
            \includegraphics[width=0.6\textwidth]{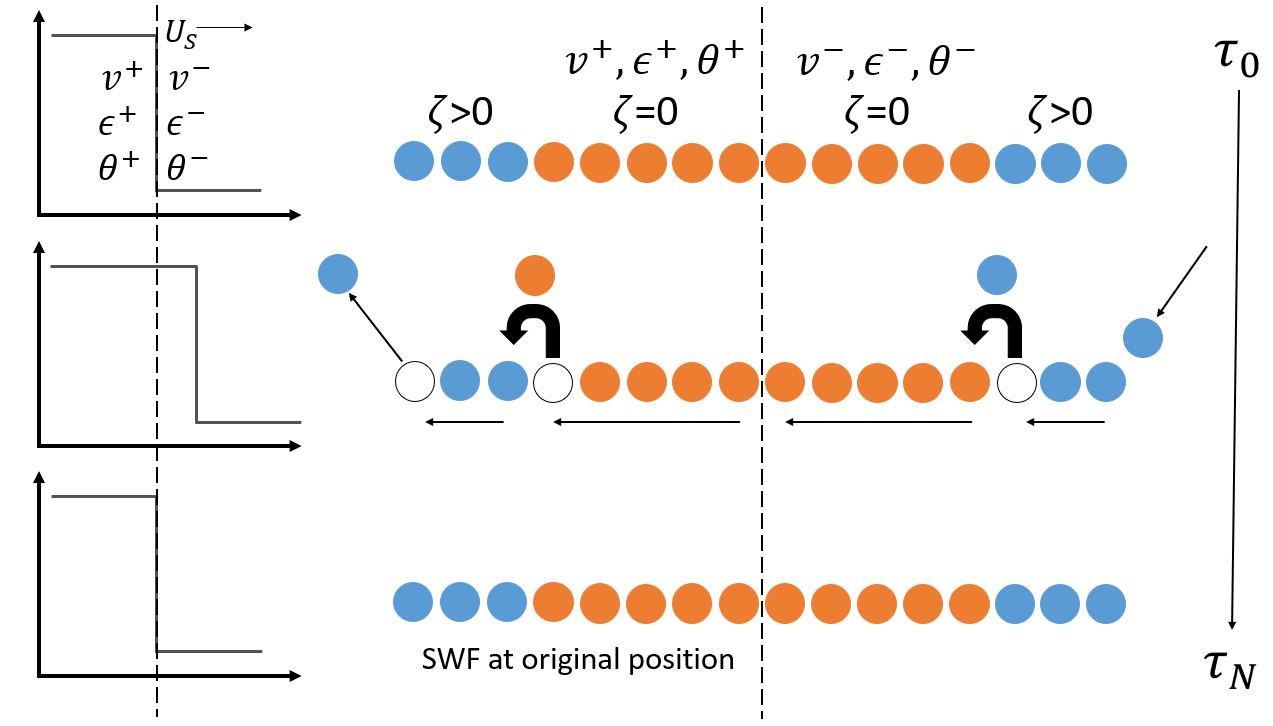}
            \caption{\textit{Schematic of the moving window mechanism for a shock wave propagating through the 1D chain.}}
            \label{fig:1DMovingWindow}
        \end{figure}

        We introduce a modified version of this moving window method into our formulation to minimize wave reflections and follow the propagating shock.
        This will allow us to perform a detailed investigation of the shock wave jump conditions and Hugoniot EOS from an atomistic point of view.  
        Our moving window atomistic method is outlined in Fig. \ref{fig:1DMovingWindow}. 
        
        The shock wave front originates at the center of the WA region through the jump conditions and shock Hugoniot as detailed in Sec. \ref{InitializationShockWave} and immediately begins propagating forward into the unshocked material.
        The moving window formulation works as follows.
        After the shock front has traveled a distance of one equilibrium lattice spacing $r_0$, atom $N_1$ is set equal to atom $N_2$, atom $N_2$ is set equal to atom $N_3$, and so on down the chain up to and including atom $N_{Total} - 1$.
        This process effectively removes the continuum atom at the leftmost end of the chain while simultaneously shifting every atom to the position of its nearest left neighbor.
        During this shifting mechanism, the window atom at the left WA/CA boundary becomes a continuum atom, and the continuum atom at the right WA/CA boundary becomes a window atom. 
        As a result of this process, atom $N_{Total}$ is effectively removed, and hence the rightmost lattice position in the chain is vacant. 
        Therefore, we insert a new $N_{Total}$ atom into the chain with position $x_{Total} = x_{Total - 1} + r_0$ (unstrained system), velocity $v^- = 0$, and acceleration $a^- = 0$.
        Local atomic energy fluctuations induced near the right boundary by the insertion of atom $N_{Total}$ are damped by the Langevin thermostat as in \cite{zhakhovsky2011two}. 
        This shifting/insertion method constitutes the moving window formulation, and it occurs iteratively with a frequency of $\tau^{-1} = U_{S} / r_0$ as the simulation progresses.
        The moving window maintains the shock front at the center of the WA region indefinitely instead of the shock propagating forward to the right boundary.  

\begin{figure}[H]
            \centering
            \includegraphics[width=0.6\textwidth]{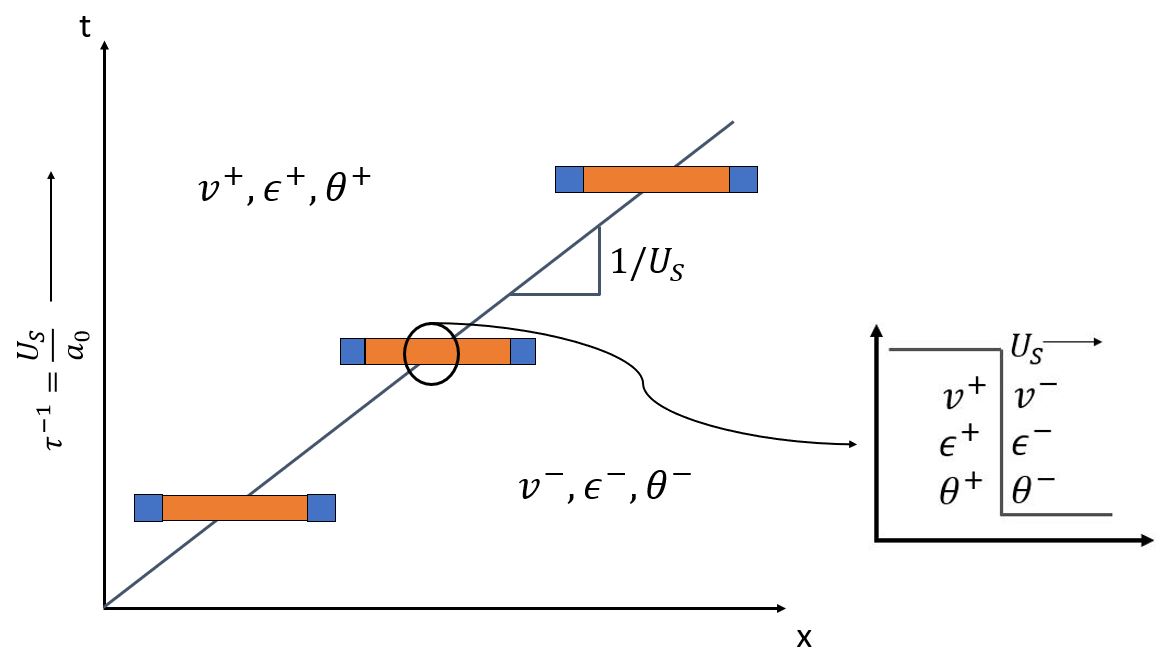}
            \caption{\textit{X-t diagram representation of the moving window formulation.}}
            \label{fig:MovingWindowXT}
        \end{figure}

        An $x - t$ diagram of the moving window method is presented in Fig. \ref{fig:MovingWindowXT}. 
        An idealized shock wave with speed $U_{S}$ originates at $(x, t) = (0, 0)$ and travels into the initially undisturbed material. 
        At time $t = 0$, the shock front is located at the center of the WA region.
        The WA regions ahead of and behind the shock are initialized with $(v^-,\epsilon^-)$ and $(v^+,\epsilon^+)$ respectively, and the CA regions ahead of and behind the shock are initialized with $(v^-,\epsilon^-,\theta^-)$ and $(v^+,\epsilon^+,\theta^+)$ respectively.
        When the simulation begins, the shock wave starts to propagate forward through the unshocked material.
        The moving window process occurs iteratively with a frequency of $\tau^{-1} = U_{S} / a_{0}$ causing the domain to essentially ``follow'' the propagating shock.
        In typical NEMD shock simulations, the shock would propagate forward and eventually encounter the domain boundary thus limiting the total simulation time. 
        In contrast, the moving window formulation allows us to perform extremely long-time shock simulations without the total number of atoms continually increasing as the simulation evolves. 

    \section{Verification} \label{Verification}
        We perform three sets of verifications to ensure that (i) the Langevin thermostat maintains a desired equilibrium temperature, (ii) the potential functions accurately represent mechanical properties, and (iii) there are no spurious reflections and artifact waves at the WA/CA interfaces.
        For the sake of brevity, details on (i) and (ii) are presented in \ref{NVTEnsemble} and \ref{ElasticModulusVer} respectively.
        In the first, we found that the Langevin thermostat maintained a canonical (NVT) ensemble for a range of different input temperatures.
        This effect was observed regardless of the potential function used. 
        In the second, we found that the LJ and Morse potentials gave accurate tangent moduli values for a range of input temperatures, and the EAM potential accurately represented both the cohesive energy and bulk modulus of copper at 0 K.
        Next, we perform the third ``steady state" verification using the atomistic framework described in Fig. \ref{fig:AC Framework}.
           
        Before presenting this data, we also explore the effect of $w$ (length of the CA region) and $\zeta_0$ (maximum damping) in Eq. (\ref{LangevinDamping}) on the system's ability to achieve steady state and obtain canonical temperature fluctuations.
        We varied the length of each CA region from 3 to 500 atoms and the maximum damping from 0.1 to 1.0 times the Debye frequency of copper ($\omega_D$). 
        To preserve the scope of this paper, we do not present the full extent of these simulations.
        Rather, we merely highlight the results. 
        
        First, we found that the length of the CA region needed to be at least the range of the forces ($\sim 11$ \AA\, for the applied EAM potential).
        When the CA region contained only 3 atoms ($\sim$ 7.67 \AA), energetic pulses were not properly damped, and thus traveling waves appeared in the WA region.
        There was, however, observed to be no upper limit to the stadium length as any CA region longer than $\sim 11$ \AA\, adequately damped out spurious phenomena and achieved a steady state. 
        Next, we found that an optimal value for $\zeta_{0}$ was 1/2 $\omega_D$.
        If the CA regions were too weakly-damped ($\zeta_{0}$ $\le$ 0.2 $\omega_D$), the system failed to achieve a canonical ensemble. 
        However, if the CA regions were too over-damped ($\zeta_{0}$ $\ge$ 0.9 $\omega_D$), pulses were not smoothly absorbed, and larger fluctuations occurred in the WA region.
        These results are consistent with those reported in \cite{qu2005finite}.
        Therefore, we conduct all subsequent simulations with 100 atoms in each CA region and a maximum damping value of $\zeta_0 = $ 1/2 $\omega_D$.
        We use a much larger CA region length than technically necessary for extra precaution and to ensure that all transient phenomena are properly damped over long-time simulations ($\sim$ 3 ns).
        
        To ensure that the WA/CA interfaces are not introducing spurious waves into the WA region, we prescribe the same continuum states to both CA regions. 
        In other words, the problem has now moved either to the left of the shock or the right of the shock in Figure \ref{fig:MovingWindowXT}.
        The average particle velocity of the system should remain equal to the initial input value with little to no increase in the average amplitude.
        A change in the average particle velocity would indicate that the system is not reaching equilibrium, while a large increase in amplitude would mean that energy is being artificially added to the WA region.
        Additionally, after a reasonable interval of time, no traveling waves should appear in the WA region.
        Such artifacts would mean that waves are not being smoothly absorbed into the CA regions and instead reflecting off the WA/CA interfaces, 
        The presence of these waves could also mean that the periodic boundary conditions are improperly implemented.
        \begin{figure}[H]
            \centering
            \includegraphics[width=0.8\textwidth]{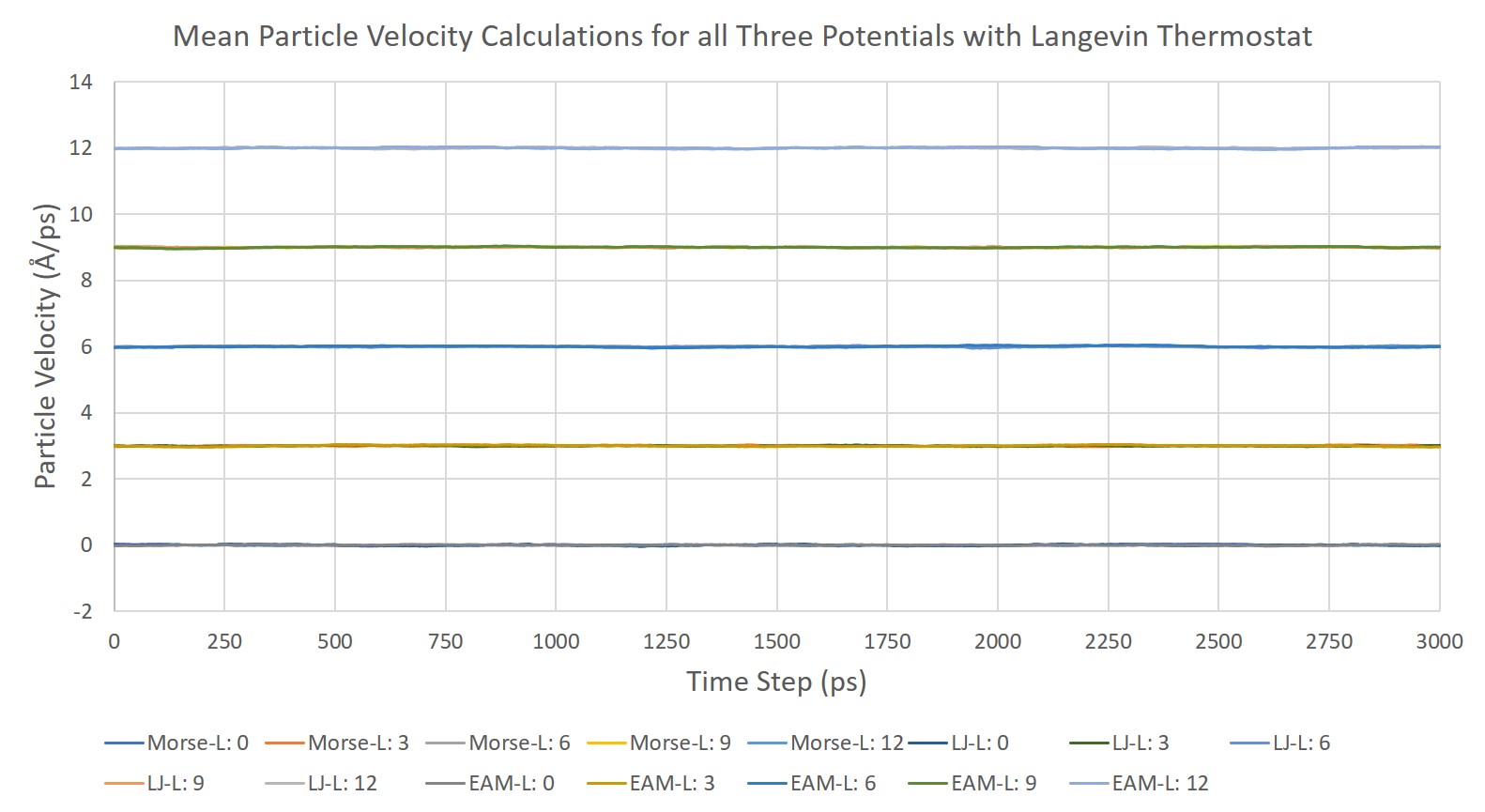}
            \caption{\textit{Average particle velocity of different systems vs. time for a range of input velocities.}}
            \label{fig:SteadyStatePlot}
        \end{figure}
            
        We perform steady state simulations for a one-dimensional chain of 10,000 copper atoms with all three potential functions.
        The Langevin thermostat is applied using the damping band method discussed in Sec. \ref{LangevinThermostat} with $\zeta_0 = 1/2 \omega_D$.
        We test the ability of the CA regions to equilibrate the system to the average input velocity and properly absorb waves / energetic pulses. 
        We perform these studies for the following mean input particle velocities: 0, 3, 6, 9, and 12 \AA/ps.
        The first set of results can be seen in Fig. \ref{fig:SteadyStatePlot}, where we plot the average particle velocity of the one-dimensional chain vs. time.
        The total run-time for each simulation is 3,000 ps (3 ns).
        We observe that the system maintains the initial mean particle velocity for the duration of the simulation for every interatomic potential.
        From these results, we conclude that the WA region achieves steady state for long-time simulations.
                 
\begin{figure}[H]
            \centering
            \begin{subfigure}{0.45\textwidth}
                \includegraphics[width=\textwidth]{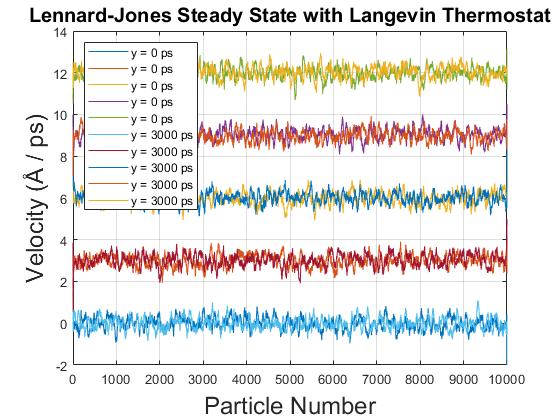}\label{fig:LJLSS}
                \caption{}
            \end{subfigure}
            \begin{subfigure}{0.45\textwidth}
                \includegraphics[width=\textwidth]{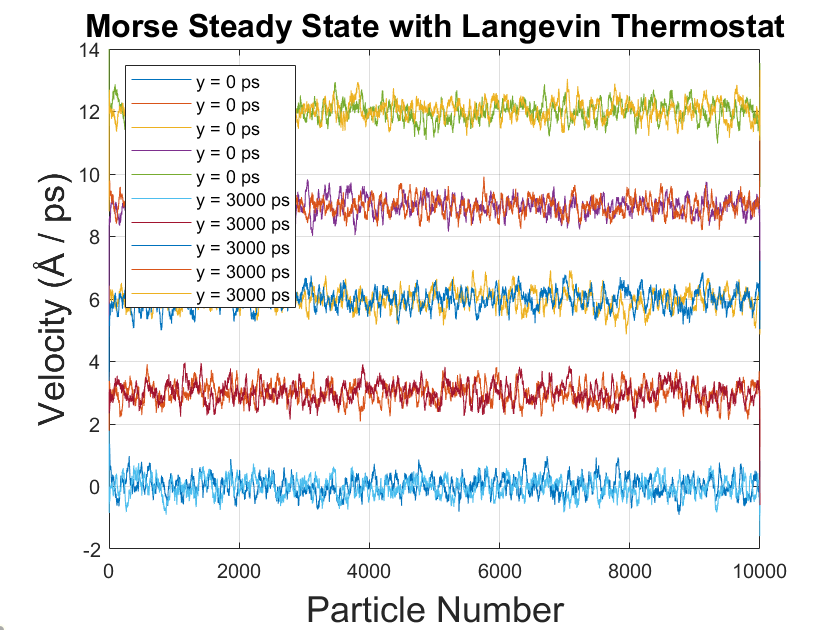}\label{fig:MLSS}
                \caption{}
            \end{subfigure}
            \begin{subfigure}{0.45\textwidth}
                \includegraphics[width=\textwidth]{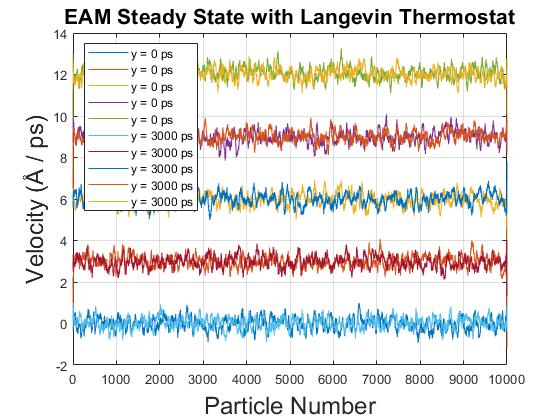}\label{fig:EAMLSS}
                \caption{}
            \end{subfigure}
            \caption{\textit{Steady state plots using the Langevin damping band method with the (a) Lennard-Jones, (b) Morse, and (c) EAM potentials.}}
            \label{fig:SteadyStateAmplitude}
        \end{figure}

        Finally, we confirm that the average amplitude of the particle velocities remains relatively constant and no large traveling waves appear throughout the duration of the simulation.
        Fig. \ref{fig:SteadyStateAmplitude} shows three different particle velocity vs. particle number plots for all three potentials.
        In each of these graphs, we plot the velocity of each particle at $0$ ps and overlay that with the velocity of each particle at $3,000$ ps for each input velocity (0, 3, 6, 9, and 12 \AA/ps).
        We overlay these sets of data from the beginning and end of the simulation to observe how the average amplitude changes over time.
        We observe that in each case, the amplitude of the particle velocities does not increase as the simulation evolves. 
        In fact, the two sets of data overlap each other almost identically indicating that no artificial energy is being introduced into the WA region.
        (We also note that no large traveling waves were observed in the WA region over the entire run-time).
        These results confirm that any waves or pulses encountering the WA/CA interfaces are traveling smoothly into the CA regions and eventually being damped out. 
        Additionally, the data establish that the periodic boundary conditions used in the CA regions are implemented correctly.

    \section{Shock Hugoniot Results}\label{ResultsEOS}
    
        \subsection{Moving window simulations using the polycrystalline EOS parameters} \label{MWShocksOldEOS}
            We perform long-time moving window shock simulations through an idealized, one-dimensional, ``close packed" chain of copper atoms. 
            As explained in Sec. \ref{ProblemStatement}, this corresponds to a planar elastic shock propagating along the [110] direction of a single-crystal copper lattice.
            The atomistic domain contains a total of 10,000 atoms with 9,800 atoms in the WA region and 100 atoms in each CA region to ensure smooth damping for long-time simulations.
            Hence, each CA region is $\frac{1}{100}$ the size of the overall domain and thus far enough away from the non-equilibrium shock wave front to be in a region of ``local" equilibrium (see Sec. \ref{InitializationShockWave} and \cite{holian1998plasticity}).
            Semi-periodic boundary conditions are enforced as described in Sec. \ref{GeomAndBCs}, the shock is initialized using the technique described in Sec. \ref{InitializationShockWave}, and the moving window is applied as detailed in Sec. \ref{MovingWindowSec}. 
            Each simulation is performed for 3,000 ps (3 ns) in order to track the motion and evolution of the fully-developed wave. 
            
\begin{figure}[H]
                \centering
                \includegraphics[width=0.6\textwidth]{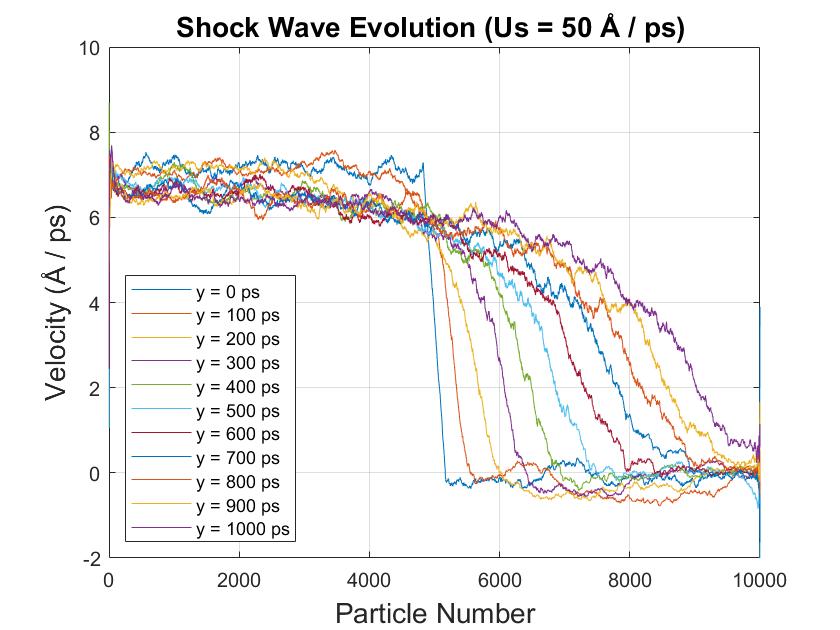}
                \caption{\textit{Propagation of a shock wave using the EAM potential for an input shock velocity of 50 \AA/ps (5.0 km/s). This simulation was produced using the shock Hugoniot parameters for polycrystalline bulk copper \cite{mitchell1991equation}.}}
                \label{fig:MWEAMLangevinOldEOS_50}
            \end{figure}

            We conduct these studies for several different shock wave velocities ($U_S$) using all three potential functions.
            However, in the following shock wave plots, we only present data obtained using the EAM potential as the other two potentials demonstrated similar phenomena. 
            We perform the first set of simulations using the Hugoniot parameters for polycrystalline copper ($C_0 = 3.94$ km/s and $S = $ 1.49 \cite{mitchell1991equation}). 
            As noted in Sec. \ref{ProblemStatement}, these values serve as an initial guess, and we will derive parameters which produce a stationary shock in Sec. \ref{NewEOSCalculations}. 
            These new EOS parameters will give us a corrected Hugoniot curve for a shock propagating along the [110] direction of a single-crystal copper lattice, and we will compare this optimized Hugoniot to those obtained in other MD studies. 
            Fig. \ref{fig:MWEAMLangevinOldEOS_50} shows a moving window shock simulation for an input shock velocity of $U_S = 50$ \AA/ps (5.0 km/s).
            In this case, we overlay the initial shock wave with its successive positions in 100 ps increments, so we see the evolution of the shock over a period of 1,000 ps. 
            
            The moving window method should maintain the shock front at the center of the WA region throughout the entire simulation.
            This is not observed in Fig. \ref{fig:MWEAMLangevinOldEOS_50}, however, because the experimental Hugoniot parameters used in our initial guess are derived for polycrystalline copper.
            Several MD studies have measured the shock Hugoniot along the different orientations of a single-crystal copper lattice and discovered large anisotropic behavior along the different crystal directions \cite{germann2000orientation,Bringa2004,lin2014effects,neogi2017shock}.
            Since we are studying a propagating shock along the [110] direction of a copper lattice, our initial Hugoniot is not suitable for this orientation.
            Such anisotropic behavior exists because plane-plane collisions propagate the shock faster along the [110] direction than along the other two directions \cite{Bringa2004,lin2014effects}.
            This causes the moving window update frequency to ``under-predict" the shock velocity causing the shock wave to drift forward towards the right boundary. 

\begin{figure}[H]
                \centering
                \includegraphics[width=0.8\textwidth]{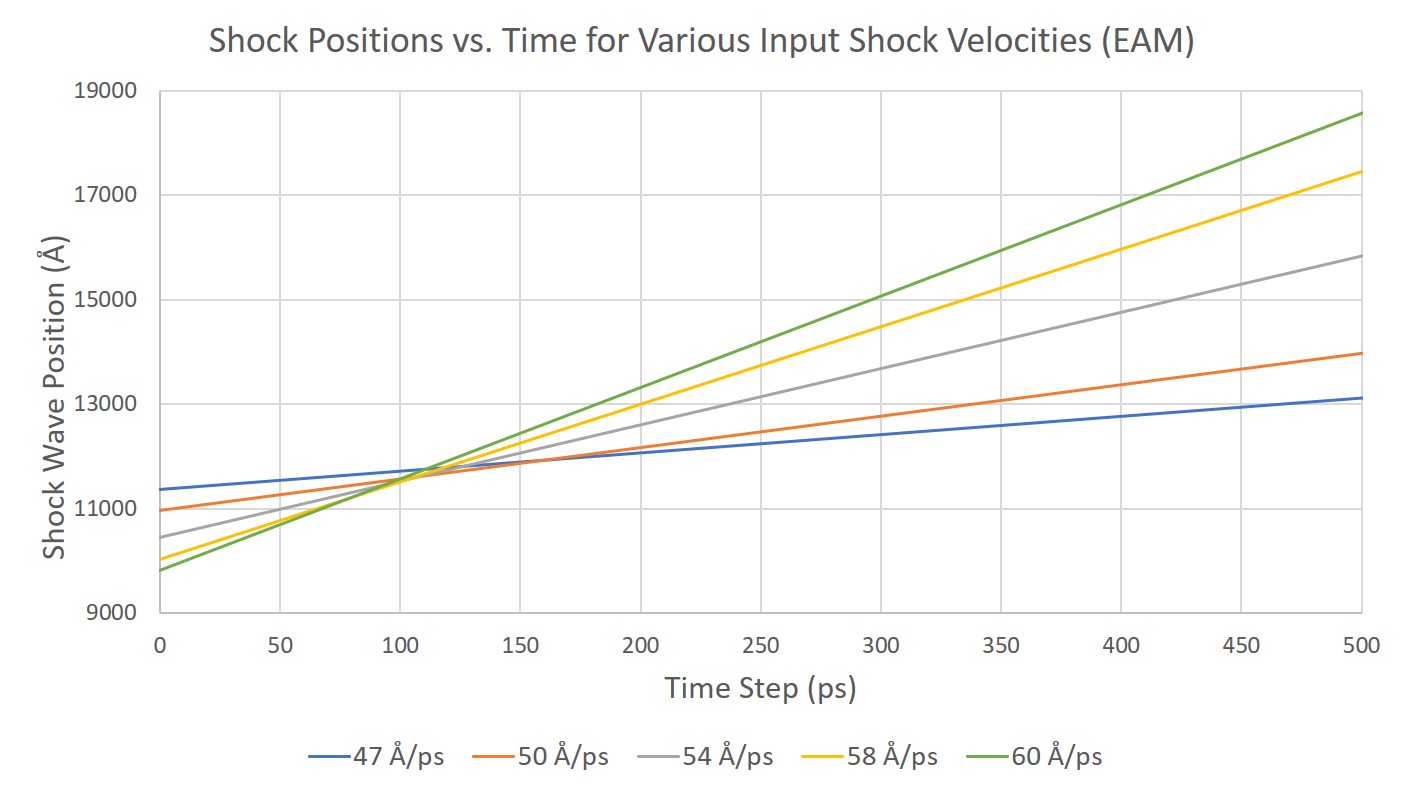}
                \caption{\textit{Position vs. time of the SWF for various input shock velocities using the EAM potential. 
                These were produced using the shock Hugoniot parameters for polycrystalline copper \cite{mitchell1991equation}.}}
                \label{fig:Old_EOS_Shock_Analysis}
            \end{figure}

            This drifting effect is also observed by plotting the shock front position vs. time for the following input shock velocities: 47, 50, 54, 58 and 60 \AA/ps.
            In Fig. \ref{fig:Old_EOS_Shock_Analysis} we observe that, in each case, the shock wave travels to the right, and the speed of this forward motion increases with increasing input shock velocity.
            (The figure terminates at 500 ps because the $60$ \AA/ps shock encountered the boundary around this time.)
            These results imply that the WA region is ``falling behind" the forward propagating shock.
            This lack of agreement between the shock wave velocity and moving window frequency becomes more pronounced as the input shock velocity increases. 
            Therefore, we must calculate the [110] shock Hugoniot by plotting the observed shock velocity vs. particle velocity directly behind the shock front.
            These results are presented in Sec. \ref{NewEOSCalculations}.
            
		\subsection{[110] shock Hugoniot calculations} \label{NewEOSCalculations}
            To derive new Hugoniot EOS parameters along the [110] direction, we analyze moving window shock simulations using all three potentials for the following input shock velocities: 47, 50, 54, 58, and 60 \AA/ps.
            We track the position of the SWF as well as the mean particle velocity behind the SWF until the shock impinges upon the right WA/CA interface (analysis after this point is invalid because the shock gets absorbed).
            To accomplish this, we fit the shock's particle velocity profile to a hyperbolic tangent function in MATLAB (using the Curve Fitting tool) for different time steps. 
            
\begin{figure}[h]
                \centering
                \includegraphics[width=0.5\textwidth]{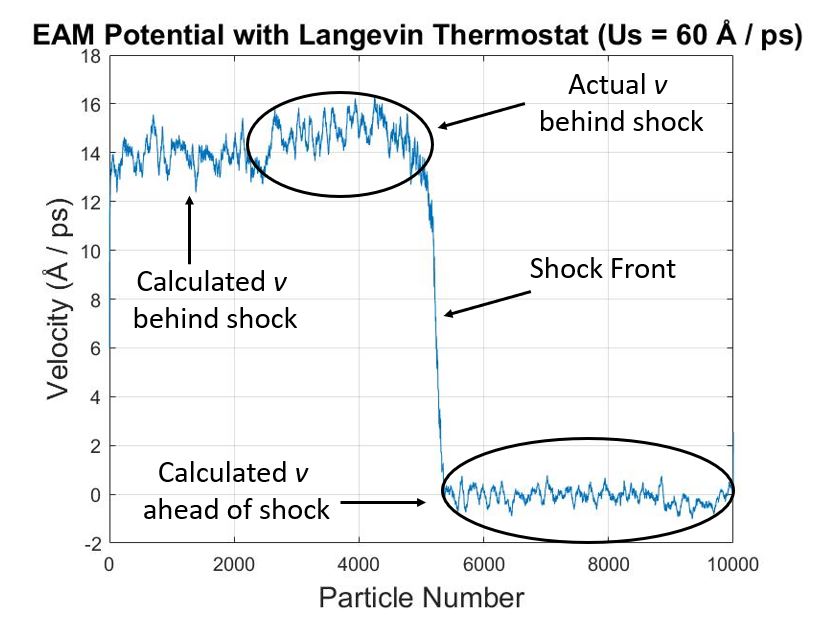}
                \caption{\textit{Snapshot at 35 ps of a propagating shock with a velocity of 60 \AA/ps (6.0 km/s).}}
                \label{fig:Shock_Wave_Analysis}
            \end{figure}

            Fig. \ref{fig:Shock_Wave_Analysis} presents a snapshot at 35 ps of a propagating shock with an input velocity of 60 \AA/ps (6.0 km/s). 
            We observe four main components in this shock profile: (i) the mean particle velocity in the shocked material ($v^+$) derived from Eq. (\ref{eq:LinearLaw}), (ii) the actual mean particle velocity behind the SWF obtained from MD, (iii) the position of the SWF (which is drifting forward), and (iv) the mean particle velocity input by the user in the unshocked material ($v^- = 0$ km/s). 
            We notice that the actual $v^+$ has a mean value which is slightly higher than that of the analytical $v^+$.
            This causes the actual $\epsilon^+$ in the shocked material to be higher than the analytical $\epsilon^+$ which results in a forward propagating shock wave.
            Therefore, because we used the polycrystalline Hugoniot parameters in our moving window simulations, the shock values ($v^+$, $\epsilon^+$, and $U_S$) obtained from MD are different from those calculated using the jump conditions.
            We use these new MD parameters to derive the shock Hugoniot along the [110] lattice direction.
            
\begin{figure}[H]
                \centering
                \begin{subfigure}{0.45\textwidth} 
                    \includegraphics[width=\textwidth]{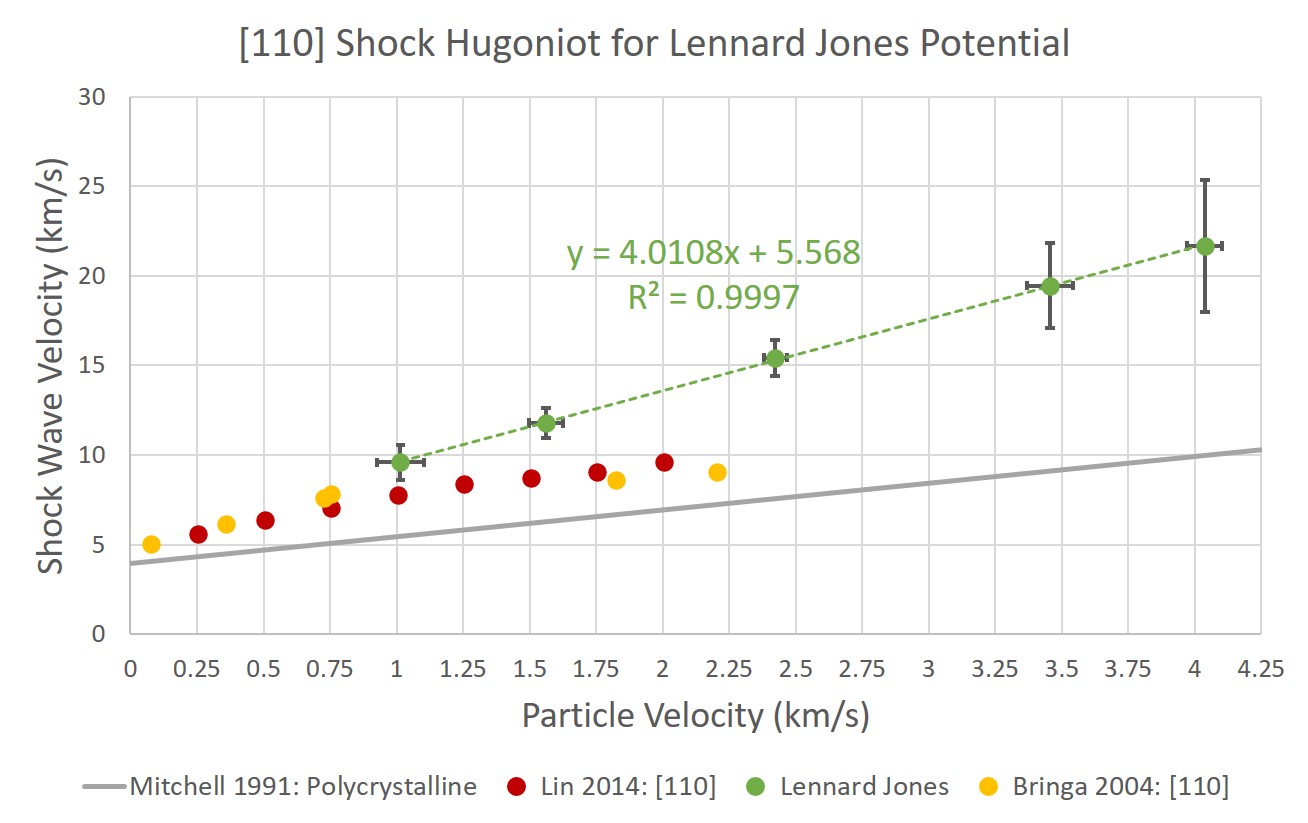} \label{110HugoniotLennardJones}
                    \caption{}
                \end{subfigure}
                \begin{subfigure}{0.45\textwidth} 
                    \includegraphics[width=\textwidth]{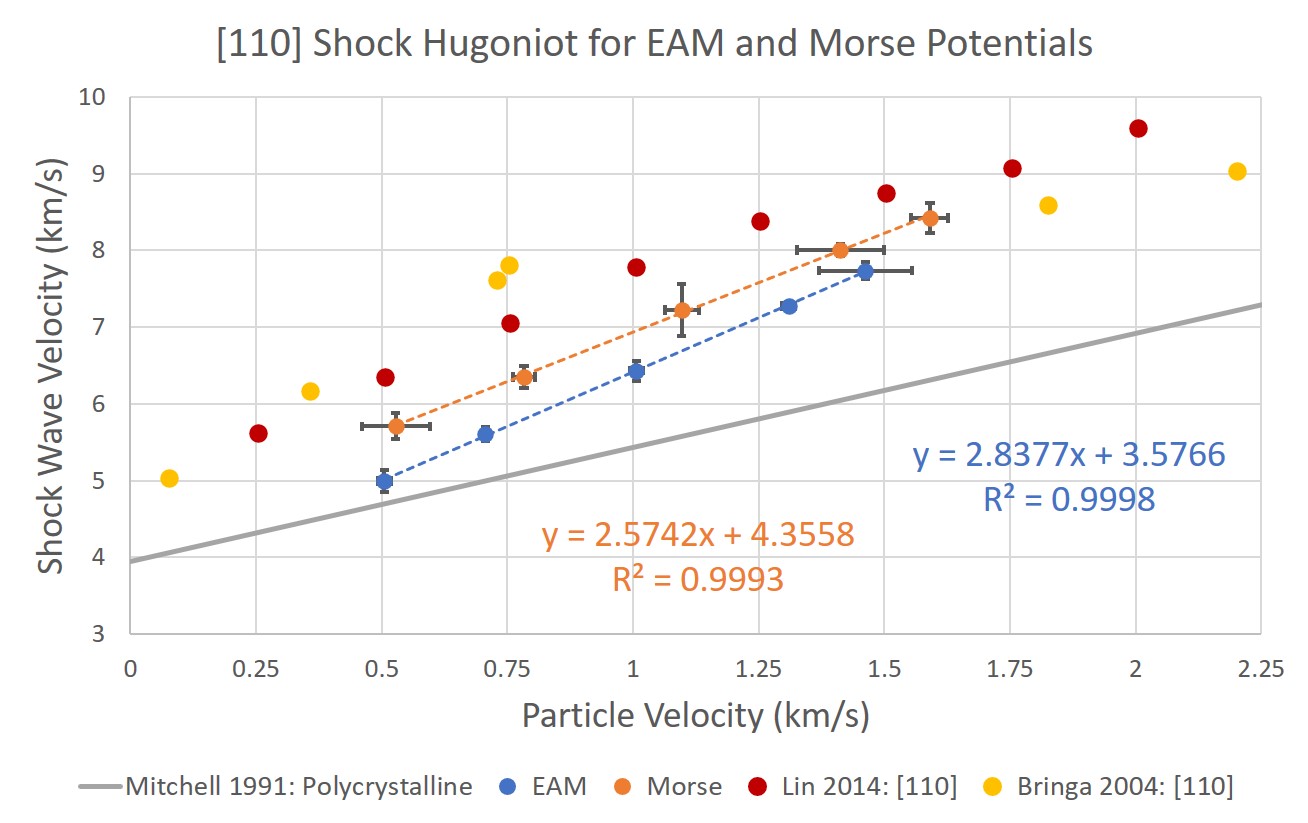} \label{110HugoniotEAMMorse}
                    \caption{}
                \end{subfigure}
                \caption{\textit{Shock velocity ($U_S$) vs. particle velocity ($v$) along the [110] crystal direction of copper using the (a) Lennard-Jones potential and (b) EAM and Morse potentials. These are compared to other NEMD simulation results for a shock along the [110] direction of a copper lattice found in \cite{Bringa2004} and \cite{lin2014effects}. Additionally, we plot experimental Hugoniot data of polycrystalline copper from \cite{mitchell1991equation}.}}
                \label{fig:NewEOSCalculationsPlot}
            \end{figure}

            Fig. \ref{fig:NewEOSCalculationsPlot} presents Hugoniot curves of the average shock velocity vs. particle velocity along the [110] lattice direction for all three potentials.
            These Hugoniots were obtained from the five shock wave trials mentioned previously, but we emphasize that the results in Fig. \ref{fig:NewEOSCalculationsPlot} are the \textit{calculated} mean shock/particle velocities from MD and \textit{not} their input values.
            The linear fits to the experimental data of polycrystalline copper by \cite{mitchell1991equation} ($U_S = 3.94 + 1.49v$) as well as NEMD shock simulation results for the [110] direction of perfect single-crystal bulk copper by \cite{Bringa2004} and \cite{lin2014effects} are also plotted in Fig. \ref{fig:NewEOSCalculationsPlot} for comparison. 
            For each potential, the slope of the linear fit is the new $S$ value while the y-intercept is the new $C_0$ value.
            
            The results from Fig. \ref{fig:NewEOSCalculationsPlot} imply that single-crystal copper is a highly anisotropic solid as the shock velocity vs. particle velocity [110] shock Hugoniots for all three potentials deviate significantly from the Hugoniot of the polycrystalline experiment. 
            Our findings corroborate the anisotropic nature of shock Hugoniots along different lattice orientations of copper observed in NEMD simulations (\cite{germann2000orientation}, \cite{Bringa2004}, and \cite{lin2014effects}) as well as MSST simulations (\cite{neogi2017shock}).
            The higher shock wave speeds along the [110] direction are due to plane-plane collisions that propagate the shock faster than along the [100] direction as mentioned previously.
            
            Our simulation results using the Morse and EAM potentials are in good agreement with both Bringa's \cite{Bringa2004} and Lin's \cite{lin2014effects} NEMD results for the low particle velocities studied with the moving window method ($v < 1.6$ km/s). 
            The slopes of the Hugoniots ($S$ values) obtained from these two potentials are very similar to the Hugoniot slopes from the two NEMD studies, but we do observe slight deviations from Bringa's results in the higher velocity region.
            This is attributed to the higher temperature in the unshocked region (298 K) employed in the current simulations \cite{peng2005pressure}. 
            Lin used an initial temperature of 300 K, which is why our moving window results agree more with the results in \cite{lin2014effects} over the full range of particle velocities studied. 
            The shock wave velocities (and thus $C_0$ values) obtained with the Morse and EAM potentials are slightly lower than the shock velocities from the NEMD studies, and this could be attributed to small transverse effects in a bulk crystal which are unaccounted for in our one-dimensional atomistic chain.
            
            The same correlation is not observed when using the LJ potential which produced shock velocities and particle velocities much higher than those in the NEMD simulations.
            These high shock speeds would result in plastic behavior which can not be captured in a one-dimensional framework.
            Additionally, the slope of the Hugoniot curve obtained from the LJ potential is much higher than the slopes from any of the other data sets. 
            Finally we note that LJ is, in general, a poor model for copper.
            Therefore, we perform all further moving window simulations using only the Morse and EAM potentials.  
            
            
            The Morse and EAM shock Hugoniot results in Fig. \ref{fig:NewEOSCalculationsPlot} are in good agreement with the NEMD results, and this provides further confirmation that a shock propagating through a one-dimensional chain of ``closed packed" copper atoms is comparable to a planar shock moving along the [110] direction of a single-crystal copper lattice.
            Additionally, these results show that the moving window formulation presented in the current paper can be used with multiple interatomic potential functions.
            We observe that EAM produces $C_0$ and $S$ values of approximately 3.577 km/s and 2.84 respectively while Morse produces values of 4.356 km/s and 2.57 respectively.
            We define these as the empirical parameters of a linear shock Hugoniot along the [110] direction of a bulk copper crystal and use them to produce a stationary shock wave in Sec. \ref{MWShockNewEOS}.
             
         \subsection{Moving window simulations with new [110] shock Hugoniot} \label{MWShockNewEOS}
\begin{figure}[H]
                \centering
                \begin{subfigure}{0.45\textwidth} 
                \includegraphics[width=\textwidth]{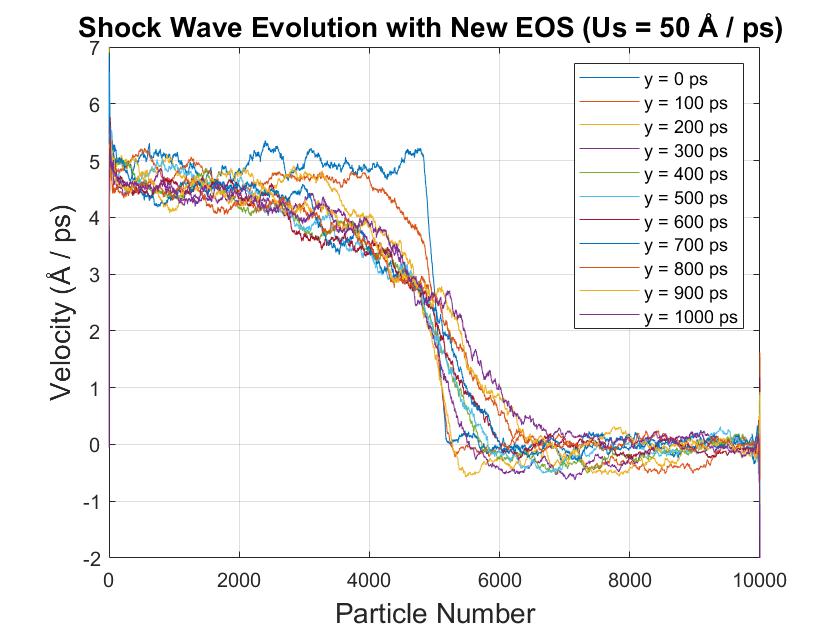}
                \caption{}
                \end{subfigure}
                \begin{subfigure}{0.45\textwidth}
                \includegraphics[width=\textwidth]{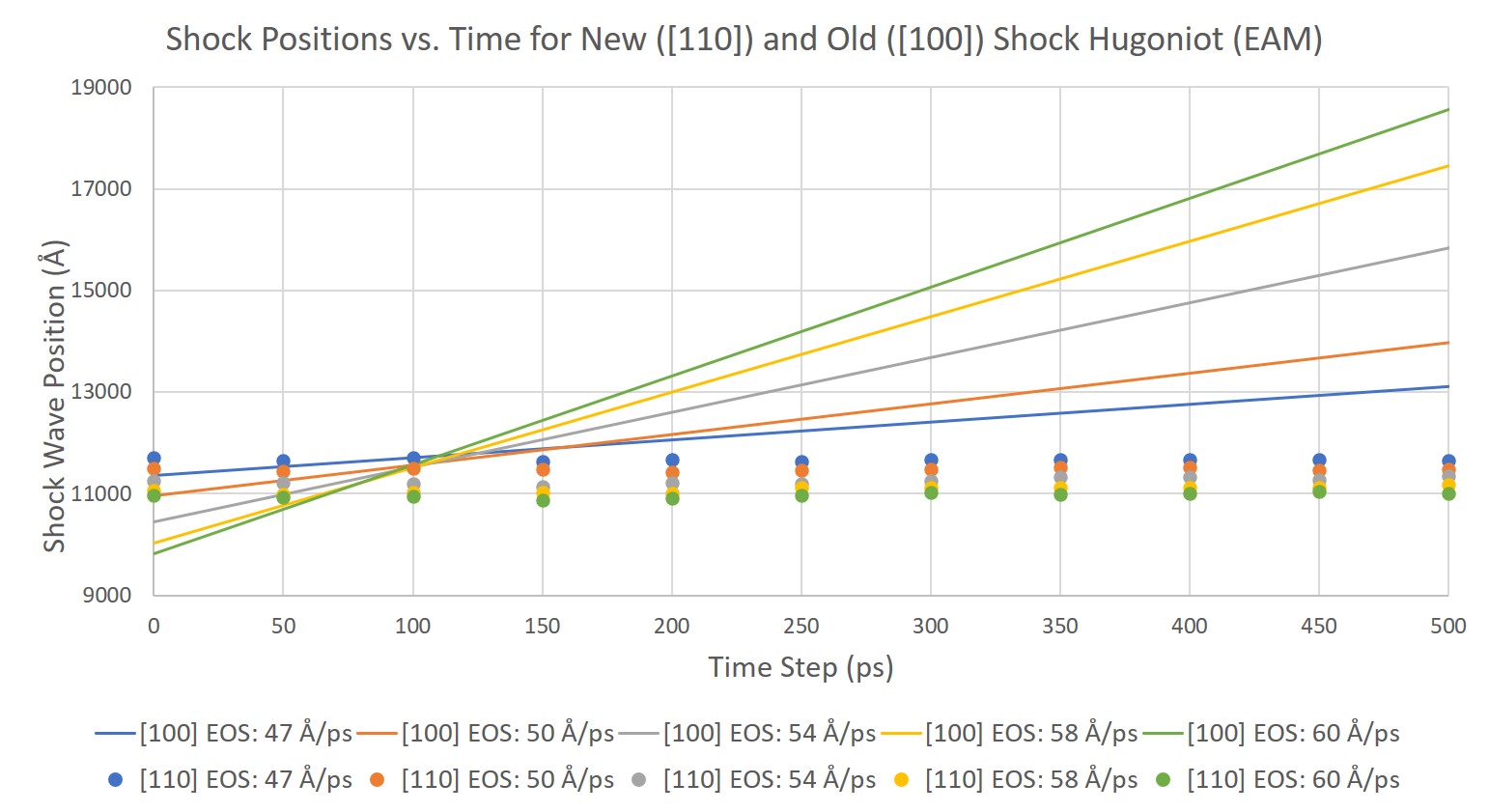}
                \caption{}
                \end{subfigure}
                \caption{\textit{(a) Propagation of a shock wave using the EAM potential and incorporating the new [110] Hugoniot EOS parameters for EAM ($U_S$ = 50 \AA /ps). (b) Shock position vs. time when using the new [110] Hugoniot compared to the results from Fig. \ref{fig:Old_EOS_Shock_Analysis}.}}
                \label{fig:MWEAMNewHugoniot}
            \end{figure}

            In Fig. \ref{fig:MWEAMNewHugoniot}a, we present the time evolution of a $50$ \AA/ps ($5.0$ km/s) shock wave over 1,000 ps in increments of 100 ps (the total run-time was 3,000 ps).
            This simulation uses the EAM potential and introduces the new [110] Hugoniot parameters for EAM into Eq. (\ref{eq:LinearLaw}). 
            We performed the same MD simulations using the Morse potential with its new Hugoniot and saw similar results, so we only present and discuss data for the EAM potential here. 
            In Fig. \ref{fig:MWEAMNewHugoniot}a, we achieve much better agreement between the input shock velocity and MD shock velocity than attained in Sec. \ref{MWShocksOldEOS}.
            The shock front is now remaining stationary in the atomistic domain and not drifting towards the right WA/CA interface.
            It is apparent that when using the new [110] shock Hugoniot parameters, the midpoint of the shock front maintains its position at the center of the WA region much longer than when using the polycrystalline Hugoniot parameters which assume that the shock propagates along the [100] lattice direction. 
            This effect is even more noticeable in Fig. \ref{fig:MWEAMNewHugoniot}b where we plot the shock position vs. time from the new Hugoniot simulations and compare these results to Fig. \ref{fig:Old_EOS_Shock_Analysis}. 
            Whereas the shock fronts were all drifting forward before, they are now remaining stationary (as evidenced by the horizontal data points) throughout the duration of each simulation.
            Therefore, the moving window update frequency now matches the shock velocity, so the atomistic domain is properly ``following" the propagating shock.
            
            For all of the shock simulations with the new Hugoniot parameters, we do observe an increase in the shock thickness over time. 
            This effect is evident in Fig. \ref{fig:MWEAMNewHugoniot}a.
            Although the midpoint of the shock front remains relatively stationary, the shock ``spreads out" across the WA region at higher time steps. 
            This is a consequence of the shock developing a structure as it propagates. 
            A structured shock wave is a well-established and characterized phenomenon \cite{chhabildas1979rise}, and it has been observed in many other one-dimensional MD shock simulations \cite{tsai1966shock,duvall1969steady,holian1978molecular,straub1979molecular,holian1995atomistic}.
            As such, care must be taken to ensure that the WA region is sufficiently large to account for the entire structured shock.
            Otherwise, the shock wave could potentially ``leak" out of the WA region and impinge on the WA/CA interface which could result in shock absorption and energy dissipation at higher time steps.
            Typically, 1D chain shock simulations result in a linear increase in the shock thickness (unsteady wave) while 3D shock simulations produce a constant shock thickness (steady wave) \cite{holian1979molecular}. 
            To understand this phenomenon further, we increase the size of the WA region and perform additional shock simulations using the new [110] Hugoniot parameters.
            These results are presented in Sec. \ref{ResultsShockWidth}.
            
    \section{Shock Structure Analysis} \label{ResultsShockWidth}
\begin{figure}[H]
            \centering
            \begin{subfigure}{0.45\textwidth}
                \includegraphics[width=\textwidth]{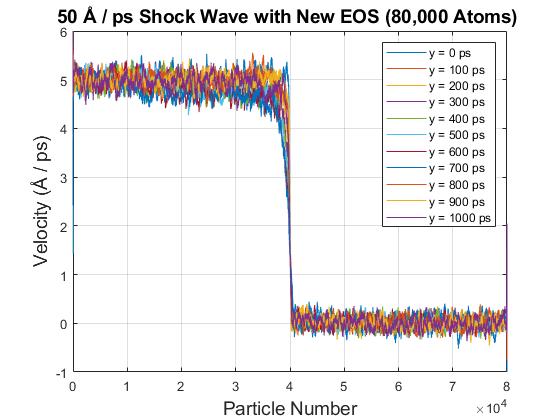}
                \caption{}
            \end{subfigure}
            \begin{subfigure}{0.45\textwidth}
                \includegraphics[width=\textwidth]{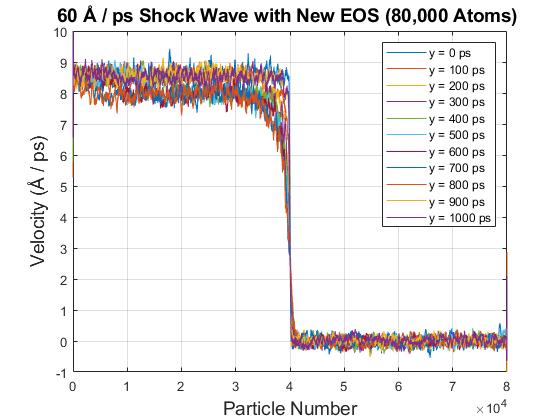}
                \caption{}
            \end{subfigure}
            \caption{\textit{Propagation of the shock wave front using the EAM potential for input shock velocities of 50 and 60 \AA / ps (5.0 and 6.0 km/s). The atomistic domain now contains 80,000 total atoms.}}
            \label{fig:NewEOS_80000_50}
        \end{figure}

        As we saw in Sec. \ref{MWShockNewEOS}, the shock wave was remaining stationary but was also developing a structure and exhibiting a length scale.
        In \cite{chhabildas1979rise}, Chhabildas and Assay calculate an upper limit of $3.0$ ns and a lower limit of $0.03$ ns for the shock rise time ($RT_S$) in copper.
        Using the new shock Hugoniot for EAM, we perform long-time moving window simulations using the following shock input velocities: 47, 50, 54, 58, and 60 \AA/ps. 
        Assuming the upper limit of $3.0$ ns for the shock rise time as well as the highest shock wave velocity of $60$ \AA/ps, we can obtain a maximum value for the shock thickness ($T_{S}$) as follows:
        \begin{equation}
            T_S = U_S \times RT_S = 60\, \text{\AA}/ps \times 3,000\, ps = 180,000\, \text{\AA}.
        \end{equation}
        Since our previous framework contained only 10,000 atoms with an equilibrium spacing of $2.556$ \AA$\,$ between atoms, the domain may not have been large enough to accommodate the shock's width. 
        This could cause part of the shock to be absorbed into the CA regions and thus dampened out during very long-time simulations ($>$ 1 ns). 
        Hence, we increase the atomistic domain size to 80,000 atoms ($\sim 204,500$ \AA) and perform shock simulations with the new [110] Hugoniot EOS.
        (Again, we only show results for the EAM potential as the Morse potential produced similar results.)
        Simulations for input shock velocities of, respectively, $50$ \AA/ps and $60$ \AA/ps can be seen in Fig. \ref{fig:NewEOS_80000_50}.
            
\begin{figure}[ht]
            \centering
            \includegraphics[width=0.7\textwidth]{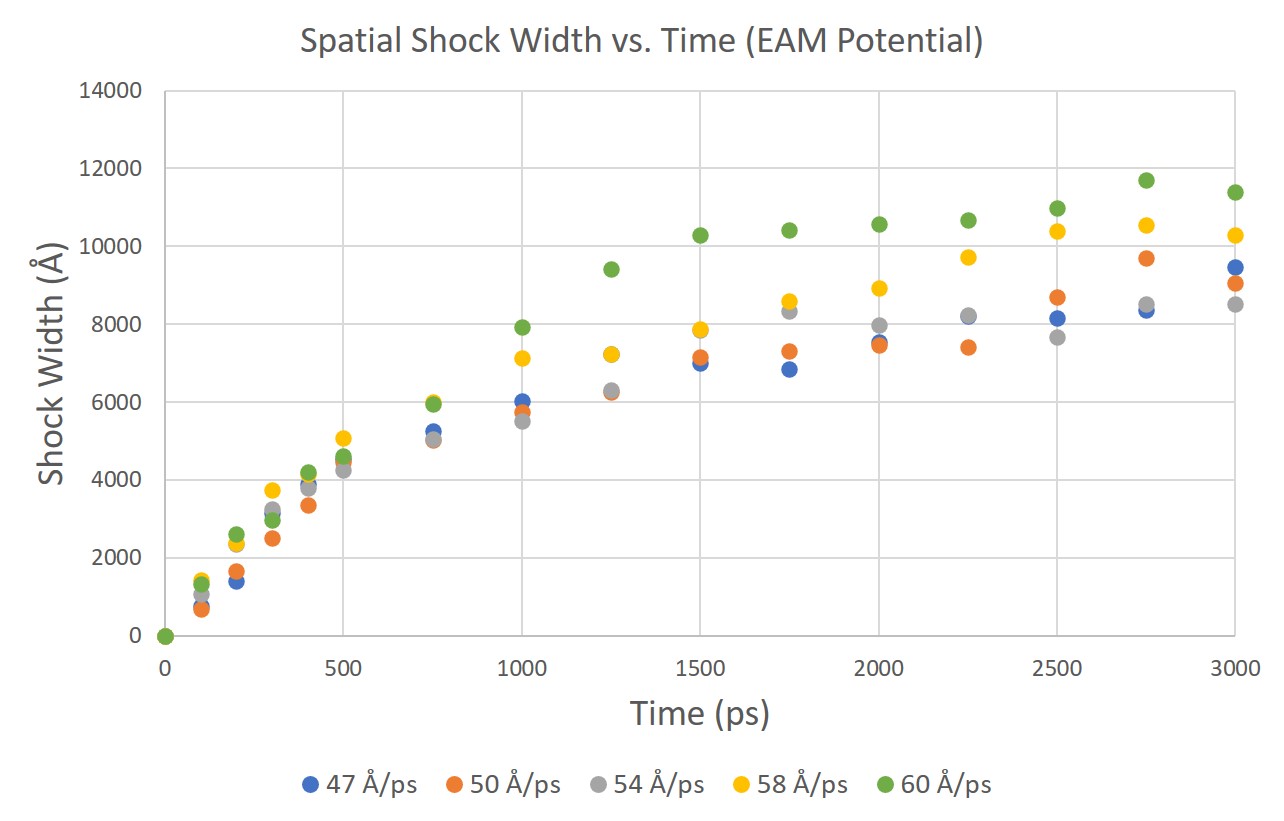}
            \caption{\textit{Spatial shock width vs. time for five different shock wave trials with new [110] Hugoniot. The EAM potential was used for these simulations.}}
            \label{fig:ShockWidth}
        \end{figure}

        For both shock wave trials, we clearly observe the shock wave maintaining its position at the center of the WA region over time. 
        Additionally, the entire structured shock is well-contained within the WA region and thus not being absorbed and dampened by the CA regions. 
        However, as seen in Fig. \ref{fig:ShockWidth}, the shock wave thickness still increases throughout the entire run-time. 
        For all five shock wave trials, the width increases linearly from 0 to 500 ps and then continues to gradually increase from 500 ps to the end of the simulation at 3,000 ps.
        As explained in Sec. \ref{ProblemStatement}, this phenomenon has been observed in many other one-dimensional shock wave studies.
        %
        Since the shock width is seen to increase for all five input shock velocities, we conclude that the one-dimensional moving window atomistic framework produces unsteady waves in agreement with other 1D chain NEMD shock simulations \cite{tsai1966shock,duvall1969steady,manvi1969finite,manvi1969shock,holian1978molecular,holian1979molecular,straub1979molecular,holian1995atomistic}.
        Our results confirm that such unsteady behavior also occurs in a ``close packed" atomistic chain assuming optimized [110] Hugoniot parameters (as opposed to experimental polycrystalline parameters) are incorporated into the linear shock Hugoniot.
            
        We note that most of the NEMD ``piston-based" shock studies which utilize a one-dimensional chain of atoms are limited to simulation times of $\le$ 100 ps because the number of atoms that have to be included grows as the shock front recedes from the piston face \cite{zhakhovsky2011two}. 
        In such simulations, a linear increase in the spatial width of the shock front is observed, and we also observe this linear growth of the shock thickness up to 500 ps in the current moving window framework.
        However, as seen in Fig. \ref{fig:ShockWidth}, the shock width growth of the five trials begins to diverge after this point. 
        This change in growth rate was not observed in previous one-dimensional NEMD simulations due to limited computational times, and such a phenomenon could be attributed to the minimal transverse motion which occurs for a shock propagating along the [110] lattice direction \cite{germann2000orientation}.
        Nonetheless, such a change in the rate of increase of the shock width is an interesting result of long-time simulations and could be a topic of future study. 
        
    \section{Conclusion}
        In this paper, we developed a moving window framework using MD to model long-time shock wave propagation through a one-dimensional chain of copper atoms.
        The framework is composed of a window region containing the shock wave flanked by boundary or ``continuum" regions on either end of the domain.
        The dynamics of the window region are governed by the classic MD equations of motion while continuum shock conditions are incorporated into the boundary regions.
        The boundary regions utilize the Langevin thermostat along with a linear damping technique to prevent spurious reflections and absorb any artifact waves that impinge on the WA/CA interfaces.
        The moving window focuses the shock wave front at the center of the WA region by adding/removing atoms to/from the WA and CA regions.
        This allows us to use relatively small domain sizes to model shock wave propagation much longer than conventional NEMD shock simulations.
        In the first part of the paper, we introduced a classic single-wave Riemann problem and defined the one-dimensional scheme.
        We then discussed the Langevin damping band method as well as the moving window formulation, and we extensively verified that each component of the framework was functioning properly.
        
        In the second part of the paper, we used the moving window framework to follow the propagating shock wave and calculate the shock velocity vs. particle velocity Hugoniot along the [110] direction of a single-crystal copper lattice. 
        We observed that the EOS parameters obtained from the Morse and EAM potentials were in good agreement with the results from other MD studies.
        We then performed moving window shock simulations with the new EOS parameters to obtain a stationary shock wave. 
        This allowed us to track the shock wave for a few nanoseconds (much longer than conventional NEMD shock simulations) and characterize the shock's structure. 
        We observed a linear increase in the width of the shock front up to 500 ps followed by a gradual increase until the end of the simulation.
        This increase in the shock front thickness was attributed to the fact that the one-dimensional framework is unable to account for plastic effects and the associated normal/transverse displacements.
        These results were consistent with early MD shock wave studies that used a one-dimensional chain of atoms.
        
        In this work, we demonstrate that the moving window formulation can follow a propagating shock wave for long simulation times ($\ge$ 1 ns). 
        Additionally, we show that a shock propagating through a ``close packed" one-dimensional chain of atoms can serve as a good approximation for a planar shock wave propagating along the [110] direction of a bulk single-crystal lattice.
        Transverse effects appear to be less influential along this lattice orientation \cite{germann2000orientation}, and our derived Hugoniot parameters are shown to be in good agreement with other MD studies \cite{Bringa2004,lin2014effects}.
        Existing NEMD shock techniques demonstrate distinct Hugoniot equations along the various lattice directions of a bulk crystal \cite{germann2000orientation,Bringa2004,lin2014effects,neogi2017shock}  while experimental methods show no such distinction \cite{chau2010shock}.
        A higher-dimensional moving window formulation could be used to resolve such a discrepancy.
        This higher-dimensional framework could also be used to follow a shock for very long simulation times and thereby study its structure, characterize its underlying kinetic relations, and examine its interactions with microstructural boundaries.
        
        The present moving window atomistic framework has the potential to be scaled up to a fully-coupled atomistic/continuum framework using methods such as CADD or CAC. 
        Such a concurrent multiscale scheme would require simultaneous refinement of the continuum region as well as coarsening of the atomistic region at the speed at which the shock wave moves.
        This would enable the atomistic region of interest to ``follow" the propagating shock indefinitely.
        While existing concurrent schemes have been very successful in modeling material defects and their motion, they have not yet been extended to model shock wave propagation through a material.
        Therefore, a multiscale scheme which can follow a propagating shock wave through the coupled domain is very much needed. 
        
    \section{Acknowledgments}
        This work was made possible by the financial support of the US Department of Defense through the National Defense Science and Engineering Graduate (NDSEG) Fellowship Program. 
        Simulations were performed using the Hopper computing cluster at Auburn University.
        Authors also thank Kaushik Bhattacharya for valuable discussion during the development of the framework.
    
    \section*{References}
    \bibliographystyle{ieeetr}
    \bibliography{1D_MW_AC_Shocks}
        
    \appendix
    
    \section{NVT Ensemble} \label{NVTEnsemble}
\begin{figure}[H]
            \centering
            \begin{subfigure}{0.45\textwidth}
                \includegraphics[width=\textwidth]{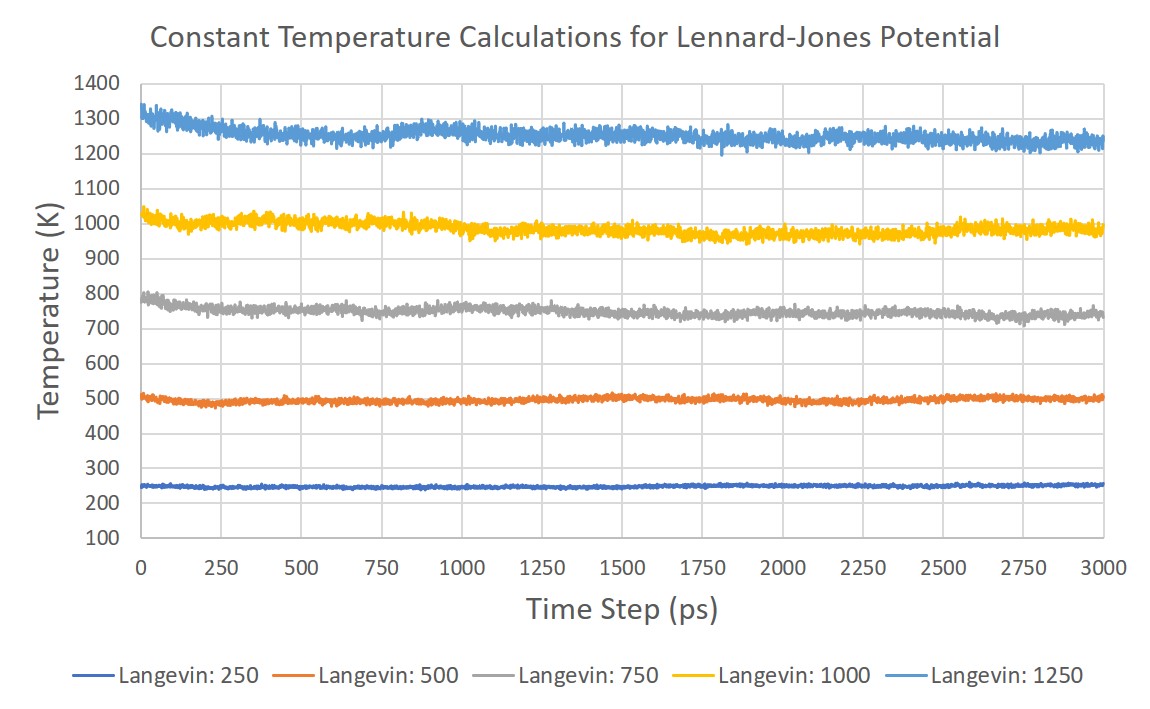}
                \caption{}
            \end{subfigure}
            \begin{subfigure}{0.45\textwidth}
                \includegraphics[width=\textwidth]{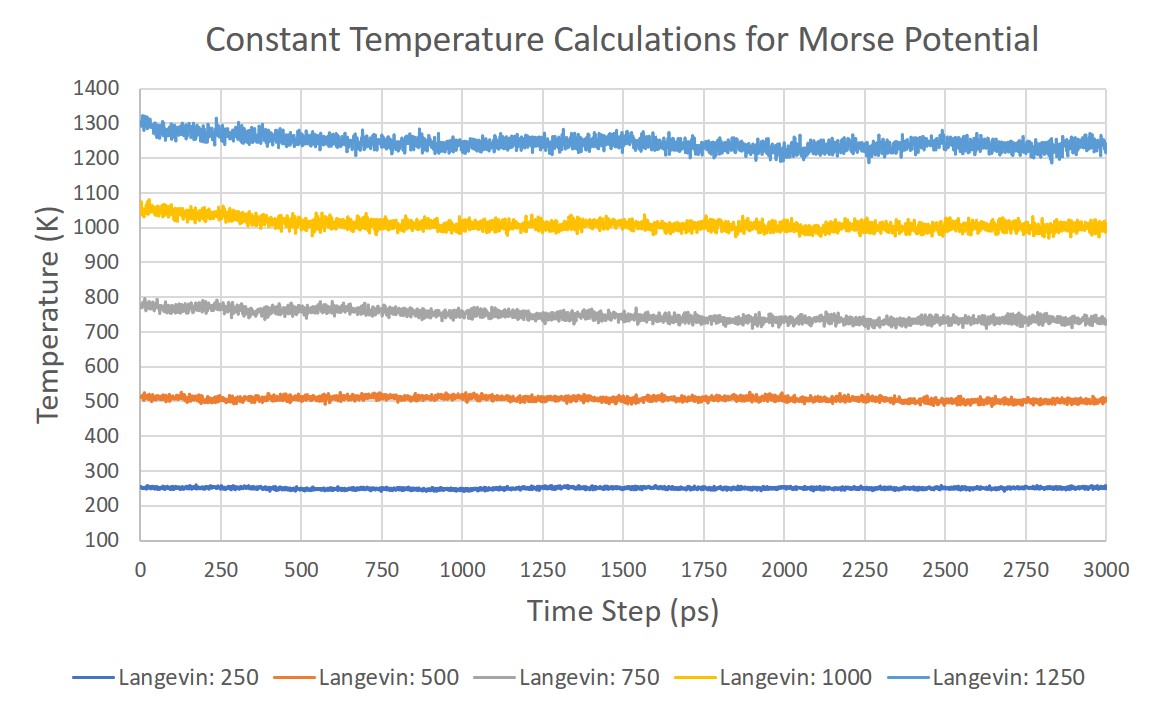}
                \caption{}
            \end{subfigure}
            \\
            \begin{subfigure}{0.5\textwidth}
                \includegraphics[width=\textwidth]{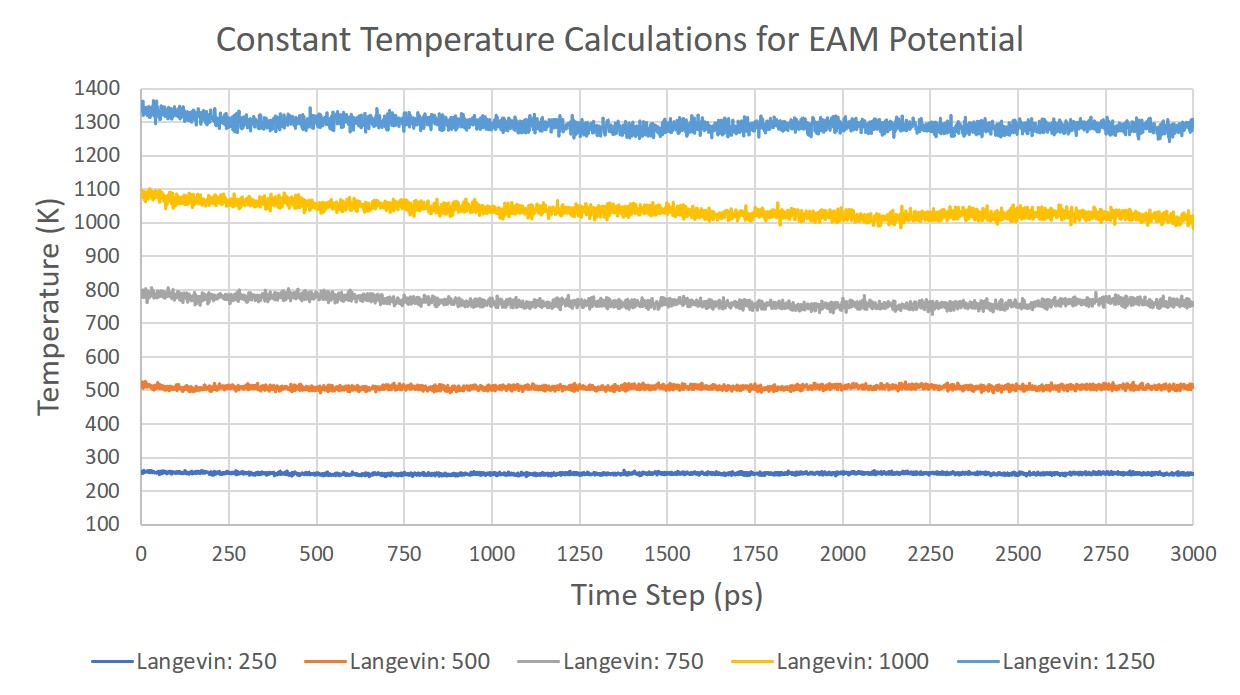}
                \caption{}
            \end{subfigure}
            \caption{\textit{Constant temperature NVT results for the Langevin thermostat using (a) Lennard Jones, (b) modified Morse and (c) EAM potentials.}}
            \label{fig:NVTPlots}
        \end{figure}

        We perform constant temperature simulations for systems of 10,000 copper atoms using the Langevin thermostat which is designed to maintain a canonical (NVT) ensemble.
        %
        We test the performance of the Langevin thermostat using all three potential functions (LJ, modified Morse, and EAM) at temperatures ranging from 250 K - 1,250 K. 
        Since the melting temperature of copper is 1,358 K, we do not perform simulations with higher input temperatures.
        The total run-time for each simulation is 3,000 ps (3 ns) with an equilibration time of 5 ps. 
        In this case, standard periodic boundary conditions are enforced such that the leftmost atom interacts with the rightmost atom in the chain and vice versa. 
        Therefore, we do not enforce the WA/CA domain described in Fig. \ref{fig:AC Framework}. 
        The results from these MD simulations can be seen in Fig. \ref{fig:NVTPlots}.
            
        For all three potential functions, the average temperatures oscillate around their corresponding initial input values for the entire run-time of 3,000 ps. 
        However, we notice that the variance in the average temperature does increases with increasing input temperature.
        This effect occurs regardless of which potential function is used.
        Such a phenomenon makes physical sense because the frequency of oscillation of the particles in a solid increases as the temperature in the solid is raised. 
        Additionally, we observe that at higher temperatures, the Langevin thermostat equilibrates the system to the initial input temperature slower than at lower temperatures. 
        This effect is seen for all three potential functions, and it is most prominent at an input temperature of 1,250 K. 
        This is understandable as the Langevin thermostat is a local thermostat, and hence there is a lack of feedback between the target temperature and input temperature.
        From these results, it is apparent that we maintain a canonical (NVT) ensemble for a wide range of input temperatures.

        \section{Mechanical Properties} \label{ElasticModulusVer}
            Verification of the Lennard-Jones and modified Morse potentials is carried out by computing the tangent modulus of the system over a range of temperatures, while verification of the EAM potential is achieved by computing the cohesive energy and bulk modulus of the system at 0 K.
            In each case, we compare the simulated mechanical properties to their corresponding literature values for copper.
            
            \subsection{LJ and Morse potentials}
                To compute the isothermal elastic modulus in 1D (tangent modulus), we utilize the microscopic elasticity tensor derived in \cite{tadmor2011modeling}. 
                The conventional expression for the microscopic elasticity tensor at a temperature $\textit{T}$ is given as follows \cite{tadmor2011modeling}:
                \begin{equation} \label{microscopicTensor1}
                    c_{ijkl} = \frac{1}{V} \left[2Nk_{B}T \left(\delta_{il}\delta_{jk} + \delta_{jl}\delta_{ik} \right) + \left<c^0_{ijkl} \right> - \frac{V^2}{k_{B}T} Cov \left(\sigma^{inst}_{ij}, \sigma^{inst}_{kl} \right)  \right]
                \end{equation}
                where $\left< \cdot \right>$ refers to a \textit{phase} average, $k_{B}$ is Boltzmann's Constant, $\textit{V}$ is the volume, and the covariance operator is defined by
                \begin{equation} \label{CovarianceOp}
                    Cov \left(A, B \right) \equiv \left<AB \right> - \left<A \right> \left<B \right>.
                \end{equation}
                Additionally, $c^0_{ijkl}$ is defined as
                \begin{equation} \label{C0Microscopic}
                    c^0_{ijkl} = \frac{1}{V} \left[ \frac{1}{4} \sum_{\alpha \ne \beta} \sum_{\gamma \ne \delta} \kappa^{\alpha \beta \gamma \delta} \frac{r^{\alpha \beta}_{i} r^{\alpha \beta}_{j} r^{\gamma \delta}_{k} r^{\gamma \delta}_{l}}{r^{\alpha \beta} r^{\gamma \delta}} - \frac{1}{2} \sum_{\alpha \ne \beta} \phi^{\alpha \beta} \frac{r^{\alpha \beta}_{i} r^{\alpha \beta}_{j} r^{\alpha \beta}_{k} r^{\alpha \beta}_{l}}{\left(r^{\alpha \beta} \right)^3} \right]
                \end{equation}
                where $\phi^{\alpha \beta}$ is the interatomic force depending only on the distance $r^{\alpha \beta}$ between the atoms and $\kappa^{\alpha \beta \gamma \delta}$ is the \textit{bond stiffness} defined by
                \begin{equation}
                    \kappa^{\alpha \beta \gamma \delta} \equiv \frac{\partial \phi^{\alpha \beta}}{\partial r^{\gamma \delta}} = \frac{\partial^2 V^{int}}{\partial r^{\alpha \beta} \partial r^{\gamma \delta}}.
                \end{equation}
                This bond stiffness is interpreted for a simple pairwise potential, where the force on atom $\alpha$ due to atom $\beta$ depends only on the distance $r^{\alpha \beta}$. 
                Equation (\ref{microscopicTensor1}) can be further simplified by splitting the instantaneous stress terms into kinetic and potential parts as seen below:
                \begin{align}
                    \sigma^{K, inst}_{ij} &= -\frac{1}{V} \sum_{\alpha} \frac{p_{i}^{\alpha} p_{j}^{\alpha}}{m^{\alpha}} \nonumber \\
                    \sigma^{V, inst}_{ij} &= \frac{1}{2V} \sum_{\alpha \ne \beta} \phi^{\alpha \beta} \frac{r_i^{\alpha \beta} r_j^{\alpha \beta}}{r^{\alpha \beta}}.
                \end{align}
                Substituting $\sigma^{inst} = \sigma^{K, inst} + \sigma^{V, inst}$ into the third term of Eq. (\ref{microscopicTensor1}) and noting that the cross-terms cancel,
                \begin{equation}
                    \left<\sigma_{ij}^{K, inst} \sigma_{ij}^{V, inst} \right> = \left< \sigma_{ij}^{K, inst} \right> \left<\sigma_{ij}^{V, inst} \right>
                \end{equation}
                we get the following:
                \begin{equation} \label{TotalCovariance}
                    Cov \left(\sigma_{ij}^{inst}, \sigma_{kl}^{inst} \right) = Cov \left(\sigma_{ij}^{K, inst}, \sigma_{kl}^{K, inst} \right) +  Cov \left(\sigma_{ij}^{V, inst}, \sigma_{kl}^{V, inst} \right).
                \end{equation}
                Then, as shown in \cite{tadmor2011modeling}, the kinetic terms can be reduced as follows:
                \begin{equation} \label{KineticCovariance}
                    Cov \left(\sigma_{ij}^{K, inst}, \sigma_{kl}^{K, inst} \right) = \left( \delta_{ik} \delta_{jl} + \delta_{il} \delta_{jk} \right) N \left(k_{B} T \right)^2.
                \end{equation}
                Substituting Eqs. (\ref{TotalCovariance}) and (\ref{KineticCovariance}) into Eq. (\ref{microscopicTensor1}), we get the simpler form of the elasticity tensor:
                \begin{equation} \label{microscopicTensor2}
                    c_{ijkl} = \frac{1}{V} \left[\left<c^0_{ijkl} \right> - \frac{V^2}{k_{B}T} Cov \left(\sigma^{V, inst}_{ij}, \sigma^{V, inst}_{kl} \right) + Nk_{B}T \left(\delta_{ik}\delta_{jl} + \delta_{il}\delta_{jk} \right)  \right].
                \end{equation}
                Here, the first term is the elasticity at 0 K, the second term is the instantaneous potential energy, and the third term is the instantaneous kinetic energy. 
                It is noted that the third term goes to zero as $T \rightarrow 0$ K. 
                Additionally, \cite{lutsko1989generalized} showed that the fluctuation term disappears as the stress and potential terms expand. 
                In this case, $c = c^{0}$, where $c^{0}$ is given by Eq. (\ref{C0Microscopic}). 
                We note that the elastic constants associated with shear vanish in the thermodynamic limit, as shown in \cite{bavaud1986statistical}. 
                However, Eq. (\ref{microscopicTensor2}) still allows us to calculate the elastic constants of solids by replacing the phase averages with time averages. 
            
                The method just described to calculate the spatial elastic modulus is known as the stress fluctuation method \cite{ray1986calculation, ray1988elastic, gao2006elastic}. 
                We use this stress fluctuation method to calculate the microscopic elastic (tangent) modulus of a one-dimensional chain of copper atoms with constant length \textit{L} and constant temperature \textit{T}. 
                For the 1D case, Eq. (\ref{microscopicTensor2}) reduces to the following \cite{wen2015interpolation}:
                \begin{equation} \label{microscopicTensor1D}
                    c = \frac{1}{L}\left[2Nk_BT + L\left<c^0\right> - \frac{L^2}{k_BT}Cov\left(\sigma^{V,inst},\sigma^{V,inst}\right)\right]
                \end{equation}
                where \textit{L} is the chain length, and ``\textit{Cov}" is the covariance operator given by Eq. (\ref{CovarianceOp}). 
                Then, the $c^0$ Born term in 1D is 
                \begin{equation}
                    c^0 = \frac{1}{L} \sum_{i=1}^N \sum_{j=i+1}^N \left[\phi''(x_{ij})(x_{ij})^2 - \phi'(x_{ij})x_{ij}\right]
                \end{equation}
                where $x_{ij} = x_j - x_i$. 
                Finally, the potential part of the instantaneous stress in 1D is given as follows:
                \begin{equation} 
                    \sigma^{V,inst} = \frac{1}{L} \sum_{i=1}^N \sum_{j=i+1}^N \phi'(x_{ij})x_{ij}.
                \end{equation}
                We compare the tangent modulus obtained from MD to the tangent modulus obtained from the Quasi-Harmonic (QH) approximation. 
                The QH approximation for the temperature-dependent stress-free spatial tangent modulus of a 1D chain of atoms is \cite{tadmor2011modeling, wen2015interpolation}
                \begin{equation} \label{TangentModulusAnalytic}
                    c = a \left[\phi''(a) + \frac{k_BT}{2}\frac{\phi^{(4)}(a)\phi''(a) - (\phi'''(a))^2}{(\phi''(a))^2}\right]
                \end{equation}
                where $a = a(T)$ is the stress-free equilibrium lattice constant at temperature \textit{T}. 
                In this case, the temperature dependence of the equilibrium lattice constant is obtained through the following equation \cite{tadmor2011modeling}:
                \begin{equation} \label{LatticeSpacing}
                    \phi'(a) + \frac{k_BT}{2}\frac{\phi'''(a)}{\phi''(a)} = 0.
                \end{equation}
                This requires calculation of third and fourth derivatives of the potential function $\phi$, making calculations for EAM cumbersome.
                We use Eq. (\ref{TangentModulusAnalytic}) to obtain the analytic tangent modulus values for the Lennard-Jones and modified Morse potentials. 

                We utilize Eq. (\ref{microscopicTensor1D}) to calculate the microscopic tangent modulus of a one-dimensional chain of 10,000 copper atoms using the LJ and modified Morse potential functions. 
                We test the performance of each of these potentials using the Langevin thermostat at various temperatures. 
                For each of the input temperatures, we calculate the corresponding equilibrium lattice spacing using Eq. (\ref{LatticeSpacing}).
                Using these temperature-dependent lattice spacings, we can obtain the tangent modulus from MD simulations with Eq. (\ref{microscopicTensor1D}) and compare this to the value obtained analytically with Eq. (\ref{TangentModulusAnalytic}).
                \begin{figure}[H]
                    \centering
                    \begin{subfigure}{0.45\textwidth}
                        \includegraphics[width=\textwidth]{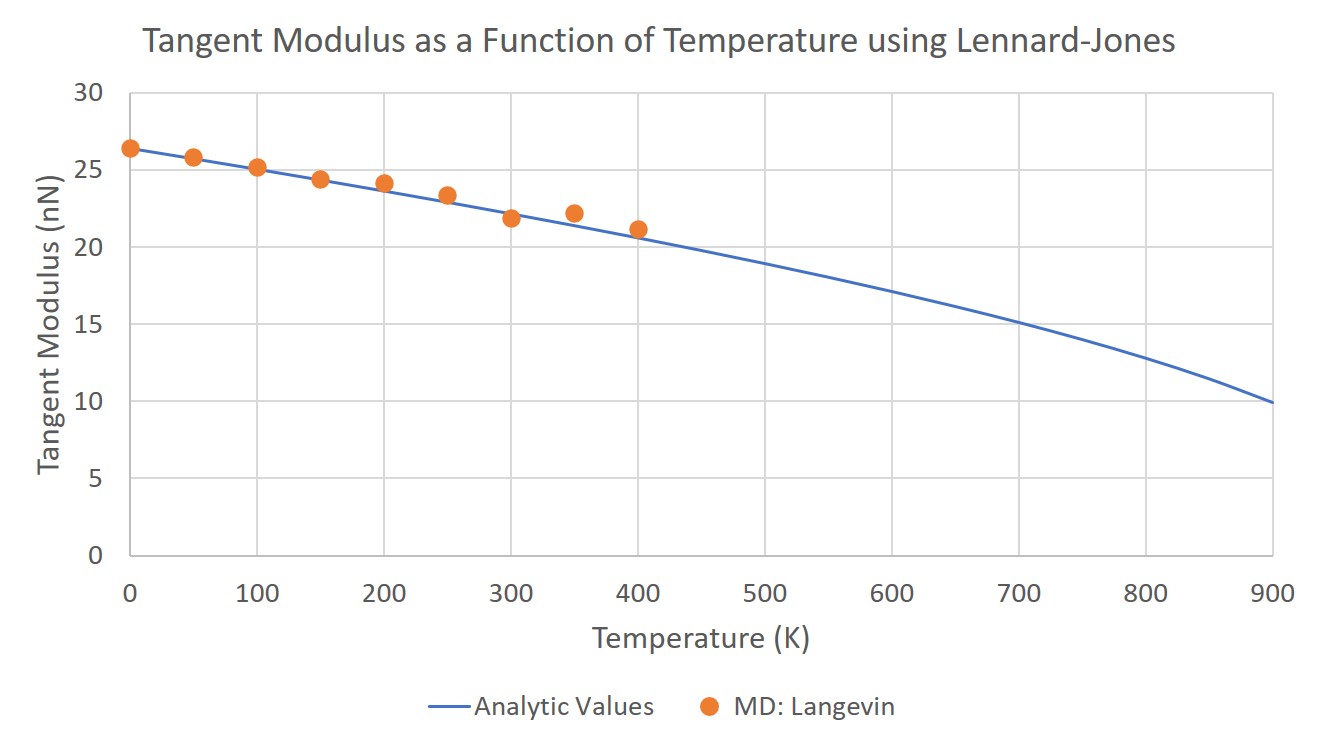}
                        \caption{}
                    \end{subfigure}
                    \begin{subfigure}{0.45\textwidth}
                        \includegraphics[width=\textwidth]{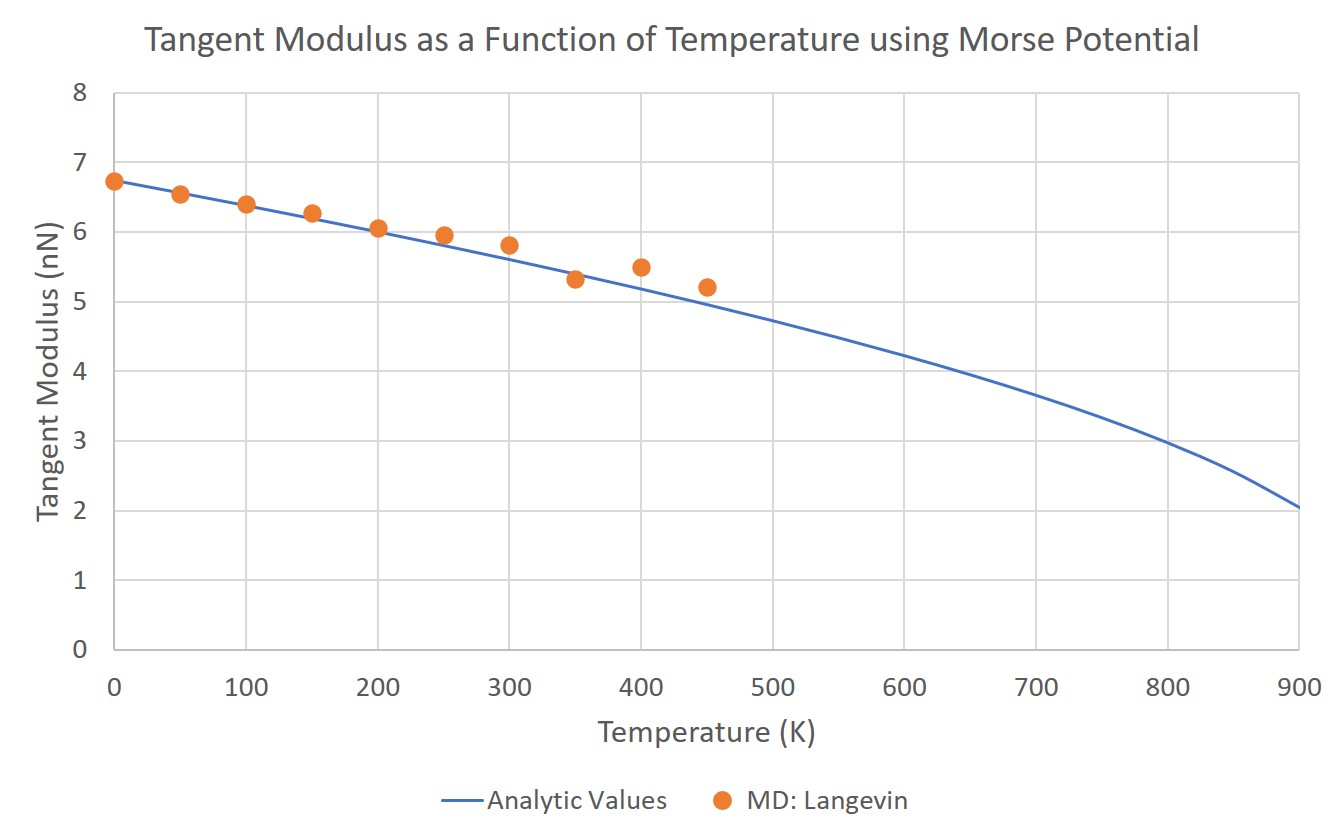}
                        \caption{}
                    \end{subfigure}
                    \caption{\textit{Tangent modulus results for the Langevin thermostat using the (a) Lennard-Jones and (b) modified Morse potentials.}}
                    \label{fig:TangentModuliPlots}
                \end{figure}
                
                Plots showing the MD and analytic tangent modulus results can be seen in Fig. \ref{fig:TangentModuliPlots}.
                Here, we present the analytic tangent modulus values (blue line) for temperatures ranging from 0 - 900 K, but we limit the MD calculations for LJ and Morse to 400 K and 450 K respectively. 
                As shown in \cite{wen2015interpolation}, the MD-derived tangent modulus of the system becomes non-physical for input temperatures above  $\approx$ 450 K. 
                The total run-time for each MD simulation is 3,000 ps (3 ns) with an equilibration time of 10 ps. 
                As in \ref{NVTEnsemble}, each atom is treated as a continuum atom, and normal periodic boundary conditions are enforced such that the leftmost atom interacts with the rightmost atom and vice versa. 
                In Fig. \ref{fig:TangentModuliPlots}, we observe that the calculated tangent modulus values from MD are in close agreement with the analytic values obtained from the QH approximation. 
                This validates the implementation of Lennard-Jones and Morse potentials in the code.
                
            \subsection{EAM potential}            
\begin{figure}[htb]
                    \centering
                    \includegraphics[width=0.6\textwidth]{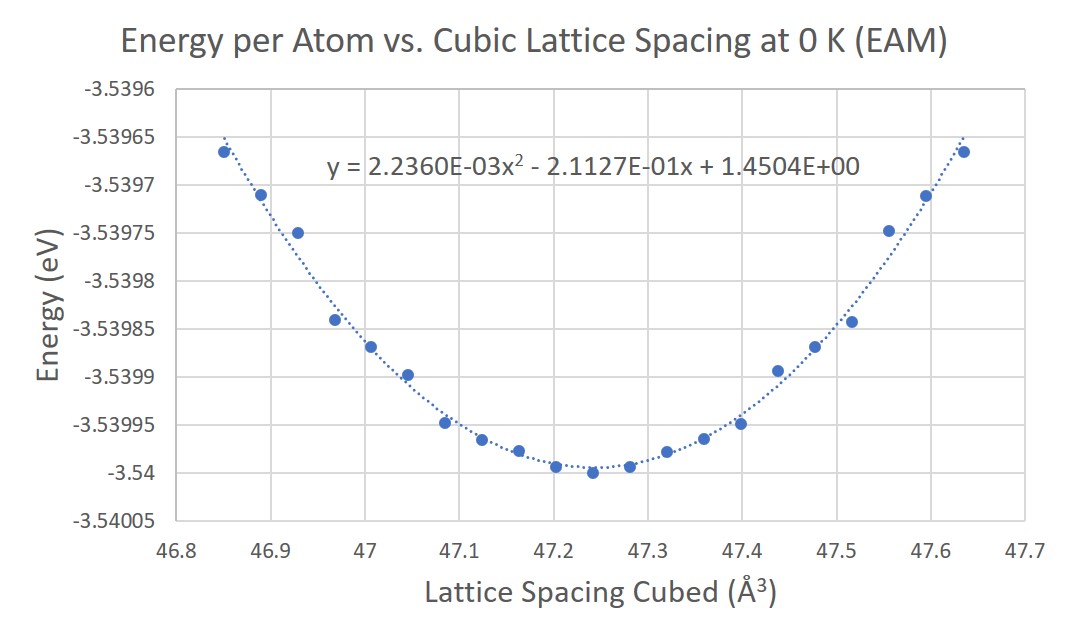}
                    \caption{\textit{Potential energy per atom vs. cubic lattice spacing in steps of $0.001$ \AA. Circles are data computed from the EAM potential, and the line is a parabola fitted to the data.}}
                    \label{fig:EAM_EnergyVsSpacing_1}
                \end{figure}

                To verify the EAM potential, we calculate the cohesive energy $E_0$ as well as the bulk modulus $B$ of the system at 0 K. 
                The cutoff radius for the EAM potential is $5.507$ \AA, and we consider a periodic chain of 500 atoms where each atom is treated as a window atom.
                The experimental value of the equilibrium lattice spacing of copper is $3.615$ \AA, so we vary the lattice constant from $3.605$ \AA\, to $3.625$ \AA\, in steps of $0.001$ \AA. 
                The potential energy per atom as a function of the cubic lattice spacing is plotted in Fig. \ref{fig:EAM_EnergyVsSpacing_1}, and the data can be fitted to a parabola.
                
                The minimum of this parabola corresponds to the cube of the equilibrium lattice spacing, $a_0 = 3.615$ \AA. 
                This matches the experimental data perfectly because $a_0$ is one of the fitted parameters of the EAM potential. 
                The energy per atom at $a_0$ is the cohesive energy, $E_{coh} = -3.540$ $eV$, which is another fitted parameter \cite{mishin2001structural}.
                Hence, our implementation of the EAM potential gives an accurate representation of the cohesive energy of copper. 
                
\begin{figure}[htb]
                    \centering
                    \includegraphics[width=0.6\textwidth]{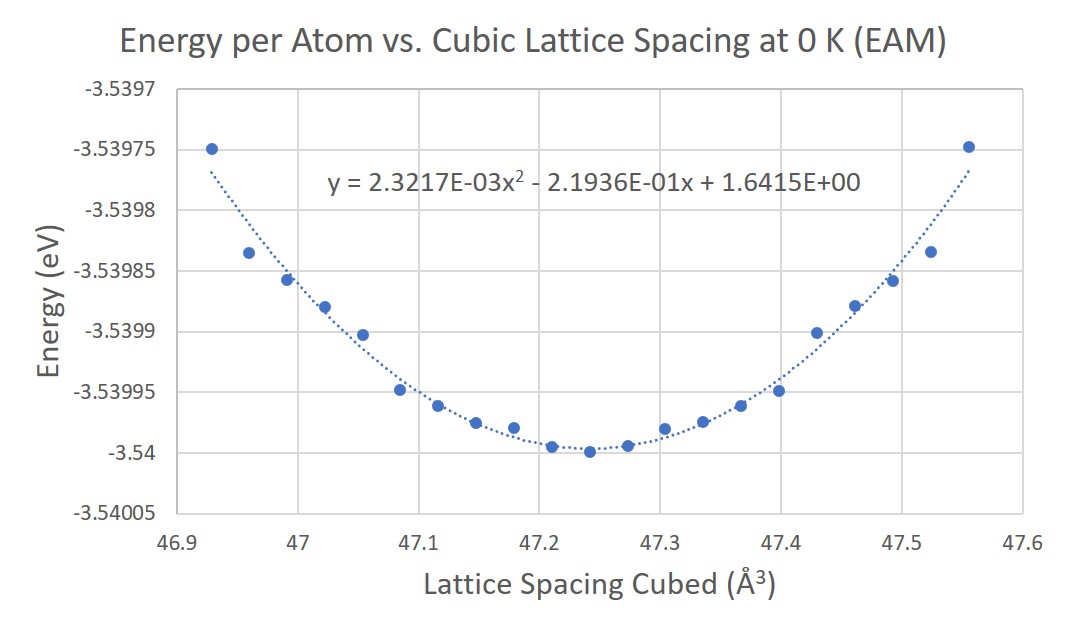}
                    \caption{\textit{Potential energy per atom vs. cubic lattice spacing in steps of $0.0008$ \AA. Circles are data computed from the EAM potential, and the line is a parabola fitted to the data.}}
                    \label{fig:EAM_EnergyVsSpacing_2}
                \end{figure}

                As discussed in \cite{cai20121}, the curvature of the parabola at $a_0$ can be used to calculate the bulk modulus using 
                \begin{equation} \label{BulkModulus}
                    B(V) = V \left(\frac{\partial^2E}{\partial{V^2}} \right)_{T,S} = 4(a_0)^3(2a)
                \end{equation}
                where \textit{a} is the parabola coefficient, and we multiply by four to account for every atom in the given lattice volume. 
                Applying this equation to the data in Fig. \ref{fig:EAM_EnergyVsSpacing_1}, we obtained a bulk modulus value of $B = 135.4$ GPa, which is not very accurate when compared to the literature value of $140$ GPa \cite{mishin2001structural}.
                To obtain a more accurate bulk modulus, we compute the $E(V)$ curve again in the range of $|a - a_0| < 10^{-4}$ \AA. 
                Specifically, we perform the calculations in steps of $0.0008$ \AA.
                This plot can be seen in Fig. \ref{fig:EAM_EnergyVsSpacing_2}.
                The curvature of this new parabola at $a_0$ gives a bulk modulus value of $B = 140.6$ GPa, which is the fitted bulk modulus of this potential model.

\end{document}